\documentclass[10pt]{article}
\usepackage{epsfig}
\usepackage{bm}

\usepackage{color}
\newcommand{\eqspa}{\hspace{5mm},\hspace{5mm}}

\def\dersec#1#2{\frac{d^2#1}{d#2^2}}
\def\derparz#1#2{\frac{\partial #1}{\partial #2}}

\newcommand{\matrdue}[4]{\left[\begin{array}{cc}{#1}&{#2}\\ 
                         {#3}&{#4}\end{array}\right]}
\newcommand{\vecdue}[2]{\left[\begin{array}{c}{#1}\\ 
                            {#2}\end{array}\right]}

\newcommand{\vv}{\mathbf{v}}
\newcommand{\nn}{\mathbf{n}}

\newcommand{\brho}{\bm{\rho}}
\newcommand{\rr}{\mathbf{r}}

\newcommand{\cc}{\mathbf{c}}

\newcommand{\q}{\mathbf{q}}

\newcommand{\ie}{{\it i.e. }}
\newcommand{\eg}{{\it e.g. }}

\begin{document}
%\large
\title{Orbit Determination\\ with Topocentric Correction:\\
Algorithms for the Next Generation Surveys}
\date{Submitted June 29, 2007}

\author{Andrea Milani, Giovanni F. Gronchi, Davide Farnocchia, \\
        Department of Mathematics, University of Pisa\\ 
        Largo Pontecorvo 5, 56127 Pisa, Italy\\
        e-mail: milani@dm.unipi.it\\
	\\
        Zoran Kne\v zevi\'c\\ 
        Astronomical Observatory\\ 
        Volgina 7, 11160 Belgrade 74, Serbia\\
	\\
        Robert Jedicke, Larry Denneau\\
        Pan-STARRS, Institute for Astronomy, University of Hawaii\\
	2680 Woodlawn Drive, Honolulu, Hawaii, 96822, USA\\
	\\
        Francesco Pierfederici\\    
        LSST Corporation\\
	4703 E. Camp Lowell Drive, Suite 253\\
	Tucson, Arizona, 85712, USA}

\maketitle

\begin{flushleft}
{\bf ABSTRACT}
\end{flushleft}

Given a set of astrometric observations assumed to belong to the same
object, the problem of orbit determination is to compute the orbit
with all the necessary tools to assess its uncertainty and
reliability. Under the conditions of the next generation surveys, with
much larger number density of observed objects, new algorithms, or at
least substantial revisions of the classical ones, are needed.
The problem has three main steps, preliminary orbit, least squares
orbit, and quality control. The classical theory of preliminary orbit
algorithms was incomplete, in that the consequences of the topocentric
correction had not been fully studied. We show that it is
possible to rigorously account for the topocentric correction,
possibly with an increase in the number of alternate preliminary orbit
solutions, without impairing the overall orbit determination
performance.
We have developed modified least squares orbit determination
algorithms, including fitting methods with a reduced number of
parameters (required when the observed arcs have small curvature),
that can be used to improve the reliability of the orbit computing
procedure.  This requires suitable control logic to pipeline the
different algorithms which we have defined and validated through
numerical simulations.
We have tested the complete procedure on two simulations with number
densities comparable to that expected from the next generation all-sky
surveys such as Pan-STARRS and LSST.
To control the problem of false identification (where observations of
different objects are incorrectly linked together) we have introduced
a quality control on the fit residuals based upon an array of metrics
and a procedure of normalization to remove duplications and
contradictions in the output.  The results confirm that large sets of
discoveries can be obtained with good quality orbits and very high
success rate losing only $0.6$ to $1.3\%$ of objects and a false
identification rate in the range $0.02$ to $0.06\%$.

\begin{flushleft}
{\bf Key Words}: Celestial Mechanics; Asteroids, Dynamics; Orbits
\end{flushleft}

% ==================================================================
\section{The Problem}
\label{s:intro}

The problem of preliminary orbit determination\footnote{Also called
Initial Orbit Determination (IOD).} is old, with very effective
solutions developed by \cite{laplace} and \cite{gauss}.  Of course the
methods of observing Solar System bodies have changed radically since
classical times and have been changing even faster recently due to
advances in digital astrometry. The question is, what needs to be
improved in the classical algorithms to handle the expected rate of
data from the next generation of all-sky surveys?  Alternatively, what
can we now use in place of the classical algorithms?
 
The issue is not one of computational resources because these grow at
the same rate as the capability of generating astrometric
data\footnote{Moore's empirical law predicts an exponential growth of
the number of elements on a chip with time and this affects the number
of pixels in a CCD and the performance of the computers used to
process astrometric data in the same way.}.  Reliability is the main
problem when handling large astronomical data sets (millions of
individual detections of Solar System objects). An algorithm failing
once when used $1,000$ times may have been considered perfectly
reliable only a few years ago but in the present situation we must
demand better performance, and even more so in the near future.

This is particularly important because of the strong correlation
between difficulties in the orbit computation and the scientific value
of the discovered object.  Main Belt Asteroids (MBA) are commonplace
and their orbits are easily computed. Only a few in a $1,000$ of the
objects (to a given limiting magnitude) are the more interesting Near
Earth Objects (NEO) while few in $100$ are the equally interesting
Trans Neptunian Objects (TNO); in both cases the computation may be
much more difficult for reasons explained later. Thus an algorithm
computing orbits for $99\%$ of the discoveries may be failing on a
large fraction of the more interesting objects like the NEO and TNOs.

To find reliable algorithms we need a good qualitative understanding
of the solutions and a firm control of the approximations used. The
classical algorithms contain several approximations, we will show that
one of them is the most dangerous: neglecting the topocentric
correction and assuming that the observer sits at the center of mass
of the Earth. The well known method of reducing to the geocenter by
assuming the object's distance and then iterating the preliminary
orbit computation does not solve the problem.

Although Laplace's and Gauss' methods each have their supporters it is
generally believed that they are equivalent to a good approximation. In
fact, we show that this is true only when the topocentric
correction is neglected. As already pointed out by \cite{marsden85},
Gauss' method has significant advantages in the way it can include the
topocentric correction. However, the classical qualitative theory
\cite{charlier} of the number of alternate solutions for the
preliminary orbit applies only to Laplace's method neglecting
the topocentric correction.

We show that a reliable method, effective under the present observing
conditions, can account for the topocentric correction in Gauss'
method although such a correction could also be included in Laplace's
method. Then we need a qualitative theory, replacing the one of
Charlier, for Gauss' method that does not assume geocentric
observations.
We have developed such a theory and found that the number of alternate
preliminary orbit solutions can be larger than in Charlier's theory,
namely, there may be double solutions at opposition and triple
solutions at low solar elongations.

A reliable orbit determination algorithm should begin with a
preliminary orbit algorithm that accounts for the possibility of
double and even triple solutions. This is especially important to
reliably handle NEO discoveries which are affected by non-unique
solutions (occurring in most cases near quadrature and sometimes even
at opposition).
Moreover, particular provisions are required for the case in which the
quantities used as input to the preliminary orbits, the curvature
components, are poorly determined to the point that their sign is
uncertain. This happens when the observed arc is too short and when
the observed object is very distant. For the case in which TNOs are
being discovered, weakly determined or totally undetermined
preliminary orbits are the rule. Two classes of methods can be used in
such cases: the \emph{Virtual Asteroids (VA)} methods and the
\emph{constrained least squares solutions}.

A Virtual Asteroid is a fully specified orbit with six orbital
elements that is compatible with the available observations but by no
means determined by them. In different types of VA methods one to
thousands (depending upon the purpose) of VA are selected either at
random or by some geometric construction inside the region in the
space of the orbits compatible with the observations\footnote{The
method of \cite{vaisala} is one way of selecting just one orbit when
there are only two observations.  The modern VA methods are
generalizations of the one by V\"aisal\"a with the advantage that they
also perform well far from opposition.}.  There are more than half a
dozen VA methods available in the literature and we will not review
all of them\footnote{For a recent review see \cite{virtast}.} but just
present one version which is particularly effective for the problem we
are discussing in this paper. It consists in choosing just one VA in
such a way that both the condition of being compatible with the
observations and having an elliptic orbit are satisfied. It is derived
from the theory of the \emph{Admissible Region} we have developed
\cite{ons1}. This is especially effective when the preliminary orbits
computed with the classical algorithms (or even modern versions) are
ineffective as happens in most cases for TNOs observed near
quadrature.

The main purpose of preliminary orbits is to be used as a first guess
for the nonlinear optimization procedure (\emph{Differential
Corrections}) that identifies the \emph{nominal orbit} fulfilling the
principle of \emph{Least Squares} (of the residuals). If the
preliminary orbits are well defined by the curvature of the observed
path, then one of them is likely to be close enough to some least
squares orbit to belong to the convergence domain of the differential
corrections; then it is sufficient to use all the preliminary orbits
as first guesses. If the preliminary orbits are poorly defined they
are likely to be so ``wrong'' that they will not lead to convergence
of the differential corrections, an inherently unstable procedure when
using a starting point far from the nominal\footnote{This property is
shared by all variants of Newton's iterative method that is
notoriously prone to chaotic behavior when used outside the
convergence domain.}.

Thus it is essential to increase the size of the convergence domain by
using modified differential corrections methods. Many of these methods
exist and most of them have one feature in common: the number of
parameters determined is less than 6. There are 4-fit methods in which
2 variables are kept fixed (at the value determined by the preliminary
orbit or by some other criterion) and the others are corrected in an
iteration converging to the minimum of the sum of squares of the
residuals restricted to a 4-dimensional submanifold of the
6-dimensional space. There are 5-fit methods in which one parameter is
fixed although it does not need to be one of the orbital elements but
might be defined in some intrinsic way adapted to the particular
problem at hand. The 5-fit method we use is fully documented in
\cite{multsol}.

Although the focus of this paper is on new theory and on algorithm
documentation we felt the need to thoroughly test the actual
performance when our methods and software are confronted with the
expected data rate of the next generation surveys. Thanks to the
collaboration with the Pan-STARRS \cite{jedicke} and LSST
\cite{ivezic} projects we have had the opportunity to use simulations
of future survey observations to precisely measure performances using
the simulation source catalog as ground truth. In particular we report
here on two tests, one based on a small but focused simulation
containing only the most difficult orbits (NEOs and TNOs), and one
representing the full data rate from a next generation survey
containing all classes of solar system objects (of course the numbers
are dominated by MBA). We believe the results are a convincing
confirmation that we are ready to handle the data from the next
generation surveys.

\section{Equations from the Classical Theory}
\label{s:classical}

There are so many different versions of preliminary orbit
determination methods and there is so little in the way of a standard
notation that it is not possible to simply copy the equations from
some reference. Thus, we have chosen to provide here a compact summary
of the basic formulae, in particular discussing the \emph{dynamical
equation} and the associated polynomial equation of degree 8 for both
methods developed by Gauss and Laplace. We also summarize Charlier's
theory on the number of solutions for Laplace's method, to be compared
to the new qualitative theory, including topocentric correction, of
Section~\ref{s:qualitative}.

\subsection{Laplace's Method}
\label{s:laplace}

The observation defines the unit vector $ \hat{\brho}
=(\cos{\delta}\cos{\alpha},\cos{\delta}\sin{\alpha},\sin{\delta}) $
where $(\alpha, \delta)$ are the topocentric right ascension and
declination.  The heliocentric position of the observed body is
\[
\rr
=\brho+\q
=\rho\hat{\brho}+q\hat{\q}
\]
where $\q$ is the observer's position\footnote{Aberration, a
consequence of the finite speed of light $c$, is accounted for by
assuming a time at the object $t_{obj}=t-\rho/c$ different from the
observation time $t$.}.  Let $s$ be the arc length parameter for the
path described by the relative position $\hat{\brho}(t)$ and $\eta$
the proper motion
\[
\frac{ds}{dt}=\eta
=\sqrt{\dot{\alpha}^2\cos^2{\delta}+\dot{\delta}^2}\ ;\ \frac{d}{ds}
=\frac{1}{\eta}\frac{d}{dt} \ .
\]
We use the moving orthonormal frame \cite[Sec. 7.1]{danby}
\begin{equation}
\hat{\brho}\ ,\ \hat{\vv}
=\frac{d\hat{\brho}}{ds}\ ,\ \hat{\nn}
=\hat{\brho}\times\hat{\vv}
\label{movingframe}
\end{equation}
and define the \emph{geodesic curvature} $\kappa$ by the equation
\begin{equation}
\frac{d\hat{\vv}}{ds}
=-\hat{\brho}+\kappa\hat{\nn}\ .
\label{defkappa}
\end{equation}
Then the relative acceleration is
\begin{equation}\label{ro 2 punti}
\frac{d^2\brho}{dt^2}
=(\ddot{\rho}-\rho\eta^2)\hat{\brho}+(\rho\dot{\eta}+2\dot{\rho}\eta)\hat{\vv}+(\rho\eta^2\kappa)\hat{\nn}
\end{equation}
and the differential equations of relative motions are 
\begin{equation}\label{equazione moto}
\frac{d^2\brho}{dt^2}
=\ddot{\rr}-\ddot{\q}
=\frac{\mu\q}{q^3}-\frac{\mu\rr}{r^3}
\end{equation}
with the following approximations: $\q=\q_{\oplus}$ coincides with the
center of mass of the Earth, the only force operating on both the
Earth and the object at $\rr$ is the gravitational attraction by the
Sun, \ie there are no planetary perturbations (not even indirect
perturbation by the Earth itself).
From (\ref{ro 2 punti})$\cdot\hat{\nn}=$(\ref{equazione
moto})$\cdot\hat{\nn}$ 
\[
\frac{d^2\brho}{dt^2}\cdot\hat{\nn}
=\rho\eta^2\kappa
=\mu\;q\;\hat{\q}\cdot\hat{\nn}\left(\frac{1}{q^3}-\frac{1}{r^3}\right) \ ,
\]
which can be presented in the form 
\begin{equation}\label{equazione dinamica}
C\frac{\rho}{q}=1-\frac{q^3}{r^3}\ \ \mbox{with}\ \
C=\frac{\eta^2\kappa q^3}{\mu \hat{\q}\cdot\hat{\nn}}
\end{equation}
referred to as the \emph{dynamical equation} in the literature on
preliminary orbits\footnote{It is, in fact, the component of the
dynamical equations along the normal to the path.}.  $C$ is a
non-dimensional quantity that can be $0$ when $r=q$ or undetermined
(of the form ${0}/{0}$) in the case that the $O(\Delta t^2)$
approximation fails, \ie the object is on an inflection point with
tangent pointing to the Sun.

Given $\rho$, to complete the initial conditions
$\dot\rho$ is solved from (\ref{ro 2
punti})$\cdot\hat{\vv}=$(\ref{equazione moto})$\cdot\hat{\vv}$
\begin{equation}\label{dinamica2}
-\mu\frac{\boldsymbol{q}\cdot\hat{\vv}}{r^3}
+\mu\frac{\boldsymbol{q}\cdot\hat{\vv}}{q^3}
=\rho\dot{\eta}+2\dot{\rho}\eta\ .
\end{equation}

\subsection{Charlier's Theory}
\label{s:charlier}

Eq.~(\ref{equazione dinamica}) is the basic formula for Laplace's
method using the solution in terms of either $\rho$ or $r$
which are not independent quantities. From the triangle formed by the
vectors $\q, \brho, \rr$ we have the \emph{geometric equation}
\begin{equation}\label{equazione geometrica}
r^2=\rho^2+2\rho q\cos{\varepsilon}+q^2
\end{equation}
where $\cos{\varepsilon} =\hat{\q}\cdot\hat{\brho}$ is 
fixed by the observation direction ($\epsilon=180^\circ-$ solar
elongation). The level curve $C=0$ is the \emph{zero circle} $r=q$.
By substituting $\rho$ solved from eq.~(\ref{equazione geometrica}) in
(\ref{equazione dinamica}), removing the square root by squaring and
multiplying by $C^2\,r^6$ ($C\neq 0$, otherwise $r=q$) we
obtain the polynomial equation
\begin{equation}\label{polinomio laplace}
P(r)
=C^2r^8-q^2r^6(1+2C\cos{\varepsilon}+C^2)+2q^5r^3(1+C\cos{\varepsilon})-q^8=0
\ .
\end{equation}
Since $P(q)=0$ there is the trivial root $r=q$, due to a singularity
in the spherical coordinates. There can be other \emph{spurious
solutions} of the polynomial equation~(\ref{polinomio laplace})
corresponding to $\rho < 0$ in eq.~(\ref{equazione dinamica}).

The qualitative theory of \cite{charlier} on the number of solutions
is obtained by analyzing these equations with elementary methods. The
sign of the coefficients of eq. (\ref{polinomio laplace}) is known:
$-(1+2C\cos{\varepsilon}+C^2)<0$ and $(1+C\cos{\varepsilon})>0$ (see
\cite{plummer}).
Thus there are $3$ changes of sign in the sequence of coefficients and
$\leq 3$ positive real roots. By extracting the factor  $(r-q)$
\[
P(r)=(r-q)\;P_1(r)\ ;\ P_1(0)=q^7\ \ ;\ \
P_1(q)=q^7\;C\;(C-3\cos{\varepsilon})\ .
\]
The number of solutions of the polynomial equation changes where
$P_1(q)$ changes sign, at $C=0 \Leftrightarrow r=q$ and at
$C-3\cos{\varepsilon}=0$. The latter condition defines the
\emph{limiting curve};
%\begin{equation}
%1-\frac{q^3}{r^3}=\frac{3}{2\,q^2}(r^2-\rho^2-q^2)\ .
%\label{limiting}
%\end{equation}
in heliocentric polar coordinates $(r,\phi)$, by using
$\rho^2=r^2+q^2-2\,r\,q\,\cos\phi$
\begin{equation}
4 - 3\frac{r}{q}\cos\phi = \frac{q^3}{r^3}\ .
\label{limcurve}
\end{equation}

\begin{figure}[h]
\centerline{\epsfig{figure=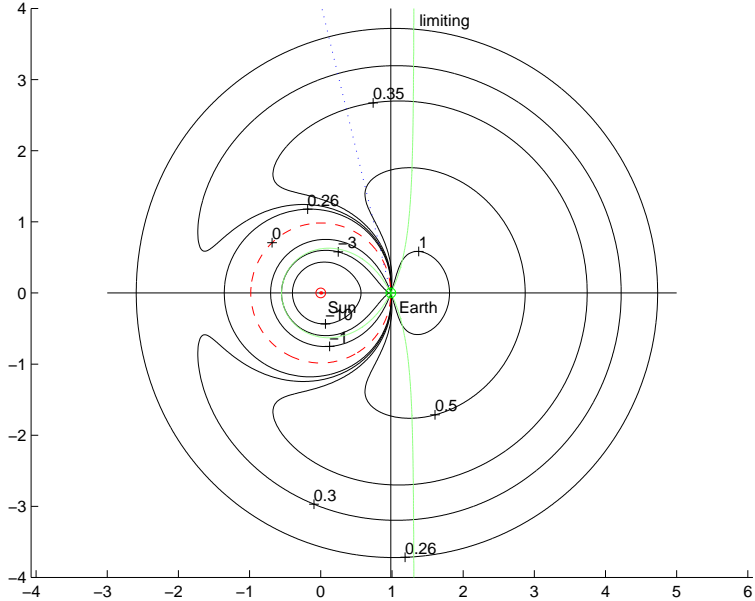,width=10cm}}
\caption{Level curves of $C(r,\rho)$ (solid lines), limiting curve
(labeled), zero circle (dashed). For a given value of $C$ and an
observation direction (dotted) there can be either 1 or 2 solutions,
\eg for $C=0.3$ there are 2.}
\label{fig:charlier2}
\end{figure}

%To understand the number of solutions we use the \emph{singular curve}
%where solutions with multiplicity 2 can occur.  By eliminating $r$
%between eq.~(\ref{equazione geometrica}) and (\ref{equazione
%dinamica}) we obtain an implicit equation connecting $C$ and $\rho$
%\begin{equation}
%F(C,\rho)
%=C\frac{\rho}{q}-1+\frac{q^3}{(q^2+\rho^2+2\,\rho\,q\,\cos\epsilon)^{3/2}}
%=0 \ .
%\label{sing1}
%\end{equation}
%At the points where $\partial F/\partial \rho =0$ the half
%line with direction defined by $\epsilon$ is tangent to some level
%curve of $C$. By using the expression of $C=C(\rho,r,q)$ and
%eq.~(\ref{equazione geometrica}) in $\partial F/\partial
%\rho =0$ we obtain the equation for the singular curve
%\begin{equation}
%1-\frac{q^3}{r^3}=\frac{3\,q^3}{2\,r^5}(r^2+\rho^2-q^2)\ ,
%\label{sing2}
%\end{equation}
%in heliocentric polar coordinates $(r,\phi)$
%\begin{equation}
%4 - 3\frac{q}{r}\cos\phi = \frac{r^3}{q^3} \ .
%\label{singcurve}
%\end{equation}
%Note that the singular curve (\ref{singcurve}) can be obtained from
%the limiting curve (\ref{limcurve}) by applying the transformation
%$r/q\mapsto q/r$.

Following \cite{charlier, charlier11} and \cite{plummer} the
number of solutions can be understood with the help of a plot of the
level curves of $C(r,\rho)$, in a plane with the Sun at $(0,0)$, the
Earth at $(q,0)$ and the position in each half-plane defined by the
bipolar coordinates $(r,\rho)$.  The limiting curve and the zero
circle can be used to deduce the number of solutions occurring at the
discovery of an object located at any point of the plane\footnote{This
plane does not correspond to a physical plane in that it also
describes the points outside the ecliptic plane.}. There is only one
solution on the right of the unlimited branches of the limiting curve,
around opposition. There are two solutions for every point of the
region between the unlimited branches and the zero circle. Inside the
zero circle and outside the loop of the limiting curve there is only
one solution. Inside that loop there are always two solutions.

Note that the classical theory by Charlier assumes that there is
always at least one preliminary orbit solution. This results from two
implicit assumptions: that the observed object exists (not being the
result of a false identification) and that the value of
$C$ is measured exactly, or at least to good accuracy, from the
observations.  $C$ contains $\kappa$ which can be difficult to
measure from a short observed arc, thus both assumptions may fail as
discussed in Section~\ref{s:distant} and \ref{s:bigtest}.

\subsection{Gauss' Method}

The method by Gauss uses 3 observations corresponding to heliocentric
positions
\begin{equation}\label{somma}
\rr_i
=\brho_i+\q_i\ \ i=1,2,3
\end{equation}
at times $t_1<t_2<t_3$ with $t_i-t_j ={\cal O}(\Delta t)\ll$ period
and the condition of coplanarity:
\begin{equation}\label{complanari}
\lambda_1\rr_1-\rr_2+\lambda_3\rr_3=0\ .
\end{equation}
From (\ref{complanari})$\times \rr_i\cdot\hat{\cc}$, where
$\cc=\rr_i\times\dot{\rr}_i$, the coefficients $\lambda_1,\,\lambda_3$
are obtained as \emph{triangle ratios}
\[
\lambda_1
=\frac{\rr_2\times\rr_3\cdot\hat{\cc}}{\rr_1\times\rr_3\cdot\hat{\cc}}
\ \ ;\ \ \lambda_3
=\frac{\rr_1\times\rr_2\cdot\hat{\cc}}{\rr_1\times\rr_3\cdot\hat{\cc}}\ .
\]
From (\ref{somma}) and
$\hat{\brho}_1\times\hat{\brho}_3\cdot$(\ref{complanari}):
\begin{equation}\label{predinamica}
\rho_2[\hat{\brho}_1\times\hat{\brho}_3\cdot\hat{\brho}_2]
=\hat{\brho}_1\times\hat{\brho}_3\cdot[\lambda_1\q_1-\q_2+\lambda_3\q_3].
\end{equation}

Next, the differences $\rr_i-\rr_2$ are expanded in powers of
$t_{ij}=t_i-t_j=O(\Delta t)$. \eg by using the $f,g$ series
formalism $\rr_i=f_i\rr_2+g_i\dot{\rr}_2\;$ and Taylor expansions
\begin{equation}\label{numeratore}
f_i=1-\frac{\mu}{2}\frac{t_{i2}^2}{r_2^3}+{\cal O}(\Delta t^3)\eqspa
g_i=t_{i2}\left(1-\frac{\mu}{6}\frac{t_{i2}^2}{r_2^3}\right)+{\cal O}(\Delta
t^4)\ .
\end{equation}
Then $\rr_i\times\rr_2=-g_i\cc$, $\rr_1\times\rr_3=(f_1g_3-f_3g_1)\cc$
and 
\begin{equation}\label{lambda}
\lambda_1=\frac{g_3}{f_1g_3-f_3g_1}>0\ ;\ 
\lambda_3=\frac{-g_1}{f_1g_3-f_3g_1}>0
\end{equation}
\begin{equation}\label{denominatore}
f_1g_3-f_3g_1
=t_{31}\left(1-\frac{\mu}{6}\frac{t_{31}^2}{r_2^3}\right)+{\cal
O}(\Delta t^4) \ .
\end{equation}
Using (\ref{numeratore}) and (\ref{denominatore}) in (\ref{lambda})
\begin{equation}\label{lambda1}
\lambda_1
=\frac{t_{32}}{t_{31}}\left[1
+\frac{\mu}{6r_2^3}(t_{31}^2-t_{32}^2)\right]+{\cal O}(\Delta t^3)\ .
\end{equation}
\begin{equation}\label{lambda3}
\lambda_3
=\frac{t_{21}}{t_{31}}\left[1
+\frac{\mu}{6r_2^3}(t_{31}^2-t_{21}^2)\right]+{\cal O}(\Delta t^3)\ .
\end{equation}

Let P be $3\times$volume of the pyramid with vertices $\q,\rr_1,\rr_2,\rr_3$ 
$P=\hat{\brho}_1\times\hat{\brho}_2\cdot\hat{\brho}_3$;
by substituting it and (\ref{lambda1}), (\ref{lambda3}) in
(\ref{predinamica}), with simple manipulations of the 
times\footnote{Use $t_{31}^2-t_{32}^2=t_{21}(t_{31}+t_{32})$ and
$t_{31}^2-t_{21}^2=t_{32}(t_{31}+t_{21})$\ .}
\begin{equation}\label{predinamica gauss}
-P\rho_2 t_{31}
=\hat{\brho}_1\times\hat{\brho}_3\cdot\left(t_{32}\q_1-t_{31}\q_2
+t_{21}\q_3\right)+
\end{equation}
\[+\hat{\brho}_1\times\hat{\brho}_3\cdot\left[\frac{\mu}{6r_2^3}[t_{32}t_{21}(t_{31}+t_{32})\q_1
+t_{32}t_{21}(t_{31}+t_{21})\q_3]\right]+{\cal O}(\Delta t^4)\ .
\] 
If the terms ${\cal O}(\Delta t^4)$ are neglected and we let
$B(\q_1,\q_3)$ represent the coefficient of the ${1}/{r_2^3}$ term in
(\ref{predinamica gauss}) then
\begin{equation}\label{B}
B(\q_1,\q_3)
=\frac{\mu}{6}t_{32}t_{21}\hat{\brho}_1
\times\hat{\brho}_3\cdot[(t_{31}+t_{32})\q_1+(t_{31}+t_{21})\q_3].
\end{equation}
Then multiply (\ref{predinamica gauss}) by
${q_2^3}/{B(\q_1,\q_3)}$ to obtain
\[
-\frac{P\;\rho_2\;t_{31}}{B(\q_1,\q_3)}\;q_2^3
=\frac{q_2^3}{r_2^3}+\frac{A(\q_1,\q_2, \q_3)}{B(\q_1,\q_3)}
\]
where
\begin{equation}\label{A}
A(\q_1,\q_2, \q_3)=q_2^3\;\hat{\brho}_1\times\hat{\brho}_3\cdot[t_{32}\q_1-t_{31}\q_2+t_{21}\q_3].
\end{equation}
Let
\[
C_0=\frac{P\;t_{31}\;q_2^4}{B(\q_1,\q_3)}\eqspa h_0
=-\frac{A(\q_1,\q_2, \q_3)}{B(\q_1,\q_3)}
\]
and then
\begin{equation}
C_0\,\frac{\rho_2}{q_2}=h_0-\frac{q_2^3}{r_2^3}
\label{dinamicagauss}
\end{equation} is the \emph{dynamical equation} of Gauss' method,
similar (but not identical) to eq.~(\ref{equazione dinamica}) of
Laplace's method.
Using (\ref{equazione geometrica}) at time $t_2$ (with $q_2$,
$\rho_2$, $r_2$ and $\varepsilon_2$):
\begin{equation}\label{polinomio gauss}
P_0(r)=C_0^2r_2^8-q_2^2r_2^6(h_0^2+2C_0h_0\cos{\varepsilon_2}
+C_0^2)+2q_2^5r_2^3(h_0+C_0\cos{\varepsilon_2})-q_2^8=0
\end{equation}
where the sign of the coefficients is as for (\ref{polinomio laplace}),
apart from $h_0+C_0\cos{\varepsilon_2}$ whose sign may change
depending upon $h_0$.  Note that $P_0(q)\neq 0$, no root
can be found analytically.  The number of positive
roots is still $\leq 3$  but a qualitative theory such as
the one of Section~\ref{s:charlier} is not available in the literature.

After the possible values for $r_2$ have been found the corresponding
$\rho_2$ values are obtained from eq.~(\ref{dinamicagauss}) and the
velocity $\dot\rr_2$ can be computed,  \eg from the classical
formulae by Gibbs \cite[Chap. 8]{herrick}.

\section{Topocentric Gauss-Laplace Methods}
\label{s:topocentric}

The critical difference between the methods of Gauss and Laplace is
the following.  Gauss uses a truncation (to order ${\cal O}(\Delta
t^2)$) in the motion $\rr(t)$ of the asteroid but the positions of the
observer (be it coincident with the center of the Earth or not) are
used in their exact values. Laplace uses a truncation to the same
order of the relative motion $\brho(t)$, thus implicitly approximating
the motion of the observer. In this section we examine the consequences
of the difference between the techniques.

\subsection{Gauss-Laplace equivalence}

To directly compare the two methods let us introduce in Gauss' method
the same approximation to order ${\cal O}(\Delta t^2)$ in the motion
of the Earth which is still assumed to coincide with the observer. The
$f$, $g$ series for Earth are
\begin{equation}\label{serie terra}
\q_i=\left(1-\frac{\mu}{2}\frac{t_{i2}^2}{q_2^3}\right)\q_2
+t_{i2}\dot{\q}_2 +\frac{\mu}{6}\frac{t_{i2}^3}{q_2^3}
\left[\frac{3(\q_2\cdot\dot{\q}_2)\q_2}{q_2^2}-\dot{\q}_2\right]+{\cal
O}(\Delta t^4)
\end{equation}
By using (\ref{serie terra}) in (\ref{B}) we find that
\[
B(\q_1,\q_3)
=\frac{\mu}{6}t_{32}t_{21}\hat{\brho}_1\times\hat{\brho}_3\cdot
[3t_{31}\q_2+t_{31}(t_{32}-t_{21})\dot{\q}_2+{\cal O}(\Delta t^3)].
\]
If $\ t_{32}-t_{21}=t_3+t_1-2t_2=0$, implying that the interpolation
for ${d^2}/{dt^2}$ is done at the central value $t_2$, then
\[
B(\q_1,\q_3)=
\frac{\mu}{2}t_{21}t_{32}t_{31}\hat{\brho}_1\times
\hat{\brho}_3\cdot\q_2\;(1+{\cal O}(\Delta t^2)) \ ;
\]
else, if $t_2\neq (t_1+t_3)/2$ the last factor is just $(1+{\cal
O}(\Delta t))$. Using (\ref{serie terra}) in (\ref{A})
\[
A(\q_1,\q_2, \q_3)=\hat{\brho}_1\times\hat{\brho}_3\cdot
\left\{-\frac{\mu}{2}t_{21}t_{32}t_{31}\q_2+\right\}
\]
%q_2^3\ \hat{\brho}_1\times\hat{\brho}_3\cdot
%\left\{(t_{32}-t_{31}+t_{21})\q_2
%+(t_{32}t_{12}+t_{21}t_{32})\dot{\q}_2\right\}-
%-q_2^3\ \hat{\brho}_1\times\hat{\brho}_3\cdot
%\left\{\frac{\mu}{2q_2^3}(t_{32}t_{12}^2+t_{21}t_{32}^2)
%+\frac{\mu}{6q_2^3}(t_{32}^3t_{12}+t_{12}^3t_{32})
%\left[\frac{3(\q_2\cdot\dot{\q}_2)\q_2}{q_2^2}-\dot{\q}_2\right]
%+{\cal O}(\Delta t^4)\right\}=
\[
%=\hat{\brho}_1\times\hat{\brho}_3\cdot\left\{-\frac{\mu}{2}t_{21}t_{32}t_{31}\q_2
\left. +\frac{\mu}{6}t_{21}t_{32}t_{31}(t_{32}-t_{21})
\left[\frac{3(\q_2\cdot\dot{\q}_2)\q_2}{q_2^2}
-\dot{\q}_2\right]+{\cal O}(\Delta t^5)\right\}.
\]
If, as above, $t_{32}+t_{12}=t_3+t_1-2t_2=0$ then
\[
A(\q_1,\q_2, \q_3)=
-\frac{\mu}{2}t_{21}t_{32}t_{31}\hat{\brho}_1\times\hat{\brho}_3\cdot
\q_2\;(1+{\cal O}(\Delta t^2))
\]
and we can conclude
\[
h_0=-\frac{A}{B}=1+{\cal O}(\Delta t^2)\ .
\]
To compute $P$ we need
\begin{equation}
\frac{d^2\hat{\brho}}{dt^2}=
\frac{d\dot{\hat{\brho}}}{dt}=
\frac{d}{dt}(\eta\hat{\vv})
=-\eta^2\hat{\brho}+\dot{\eta}\hat{\vv}+\kappa\eta^2\hat{\nn}
\label{d2rhohat}
\end{equation}
to make a Taylor expansion of $\hat{\brho}_i$ in $t_2$
\[
\hat{\brho}_i=\hat{\brho}_2+t_{i2}\eta\hat{\vv}_2
+\frac{t_{i2}^2}{2}(-\eta^2\hat{\brho}_2+\dot{\eta}\hat{\vv}_2
+\kappa\eta^2\hat{\nn}_2)+{\cal O}(\Delta t^3).
\]
This implies that
\[
\hat{\brho}_1\times\hat{\brho}_3\cdot\hat{\brho}_2=\frac 12\,
\left[t_{12}\eta\hat{\vv}_2\times t_{32}^2\kappa\,
\eta^2\hat{\nn}_2-t_{32}\eta\,\hat{\vv}_2\times 
t_{12}^2\kappa\,\eta^2\,\hat{\nn}_2\right]\cdot \hat{\brho}_2
+{\cal O}(\Delta t^5)
\]
and the ${\cal O}(\Delta t^4)$ term vanishes thus
\[
P=-\frac{\kappa\eta^3}{2}(t_{12}t_{32}^2-t_{32}t_{12}^2)\;
(1+{\cal O}(\Delta t^2))=
\frac{\kappa\eta^3}{2}t_{21}t_{32}t_{31}\;(1+{\cal O}(\Delta t^2))
\]
and
\begin{equation}
C_0=\frac{Pt_{31}q_2^4}{B}=\frac{\kappa\eta^3t_{31}q_2^4+{\cal O}(\Delta t^3)}{\mu \hat{\brho}_1\times\hat{\brho}_3\cdot\q_2\,(1+{\cal O}(\Delta t))}.
\end{equation}
In the denominator $\hat{\brho}_1\times\hat{\brho}_3$ computed to
order $\Delta t^2$ is
\begin{equation}\label{prodotto vettore}
\hat{\brho}_1\times\hat{\brho}_3
%(\hat{\brho}_2\times t_{32}\eta\hat{\vv}_2
%-\hat{\brho}_2\times t_{12}\eta\hat{\vv}_2)
%+\left(\hat{\brho}_2\times \frac{t_{32}^2}{2}\dot{\eta}\hat{\vv}_2
%-\hat{\brho}_2\times \frac{t_{12}^2}{2}\dot{\eta}\hat{\vv}_2\right)+
%+\left(\hat{\brho}_2\times \frac{t_{32}^2}{2}\kappa\eta^2\hat{\nn}_2-\hat{\brho}_2\times \frac{t_{12}^2}{2}\kappa\eta^2\hat{\nn}_2\right)=
=t_{31}\,\eta\,\hat{\nn}_2+\frac{t_{32}^2
-t_{12}^2}{2}\;(\dot{\eta}\,\hat{\nn}_2-\kappa\,\eta^2\,\hat{\vv}_2)+
{\cal O}(\Delta t^3).
\end{equation}
If  $t_{32}-t_{21}=t_3+t_1-2t_2=0$ then
\[
C_0=\frac{\kappa\,\eta^3\,t_{31}q_2^4+{\cal O}(\Delta t^3)} {\mu\,
t_{31}\,\eta\, q_2\hat{\q}_2\cdot\hat{\nn}_2 +{\cal O}(\Delta
t^3)}=\frac{\kappa\,\eta^2\,q_2^3}{\mu\,
\hat{\q}_2\cdot\hat{\nn}_2}\;(1+({\cal O}\Delta t^2)) \ ,
\]
otherwise the last factor is $(1+{\cal O}(\Delta t))$.

We can conclude that if the topocentric correction is neglected the
coefficients of the two dynamical equations (\ref{equazione dinamica})
and (\ref{dinamicagauss}) are the same to zero order in $\Delta t$
and also to order 1 if the time $t_2$ is the average
time\footnote{This equivalence is taken for granted by many authors
but only in \cite{poincare} we have found the basic idea of
the computations above.}.

\subsection{Topocentric Correction in Laplace's Method}
% =============================================================================

Now let us remove the approximation that the observer sits at the
center of the Earth and introduce the \emph{topocentric correction}
into Laplace's method.  The center of mass of the Earth is at
$\q_{\oplus}$ but the observer is at $\q=\q_{\oplus}+\mathbf{P}$.  Let
us derive the dynamical equation by also taking into account the
acceleration contained in the geocentric position of the observer
$\mathbf{P}(t)$ such that
\[
\dersec\brho t=
-\frac{\mu\rr}{r^3}+\frac{\mu\q_{\oplus}}{q_{\oplus}^3}-\ddot{\mathbf{P}}.
\]
Multiplying by $\cdot\hat{\nn}$ and using eq.~(\ref{ro 2 punti})
\[
\frac{d^2\brho}{dt^2}\cdot\hat{\nn}
=\rho\eta^2\kappa
=\mu\left[q_{\oplus}\frac{\hat{\q}_{\oplus}\cdot\hat{\nn}}{q_{\oplus}^3}
-q_{\oplus}\frac{\hat{\q}_{\oplus}\cdot\hat{\nn}}{r^3}
-P\,\frac{\hat{\mathbf{P}}\cdot\hat{\nn}}{r^3}\right]
-\ddot{\mathbf{P}}\cdot\hat{\nn}
\]
The term $P\,{\hat{\mathbf{P}}\cdot\hat{\nn}}/{r^3}$ can be neglected.
This approximation is legitimate because $P/q_\oplus\leq 4.3\times
10^{-5}$ and the neglected term is smaller than the planetary
perturbations. Thus
\begin{equation}\label{dinamica topocentrica}
C\frac{\rho}{q_{\oplus}}=(1-\Lambda_n)-\frac{q_{\oplus}^3}{r^3}
\end{equation}
where
\begin{equation}
C=\frac{\eta^2\kappa q_{\oplus}^3}{\mu \hat{\q}_{\oplus}\cdot\hat{\nn}} 
\eqspa
\Lambda_n=\frac{q_{\oplus}^2\ddot{\mathbf{P}}\cdot\hat{\nn}}
{\mu \hat{\q}_{\oplus}\cdot\hat{\nn}}=
\frac{\ddot{\mathbf{P}}\cdot\hat{\nn}}
{({\mu}/{q_{\oplus}^2})\,\hat{\q}_{\oplus}\cdot\hat{\nn}}.
\label{chi}
\end{equation}
Note that $\Lambda_n$ is singular only where $C$ is also singular. The
analog of eq.~(\ref{dinamica2}), again neglecting ${\cal
O}(p/q_\oplus)$, is
\begin{equation}
\rho\dot{\eta}+2\dot{\rho}\eta= \frac{\mu\; \hat{\q}_\oplus\cdot
\hat\vv}{q_\oplus^2}\; \left(1-\Lambda_v-
\frac{q_\oplus^3}{r^3}\right)\eqspa \Lambda_v=\frac{q_\oplus^2\;
\ddot{\mathbf{P}}\cdot \hat\vv}{\mu \; \hat{\q}_\oplus\cdot \hat\vv} \ .
\label{chi_v}
\end{equation}
The important fact is that $\Lambda_n$ and $\Lambda_v$ are by no means
small. The centripetal acceleration of the observer (towards the
rotation axis of the Earth) has size $\Omega_\oplus^2\,R_\oplus\,
\cos\theta$ where $\Omega_\oplus$ is the angular velocity of the
Earth's rotation, $R_\oplus$ the radius of the Earth and $\theta$ the
latitude; the maximum of $\simeq 3.4\; cm\,s^{-2}$ occurs at the
equator.  The quantity ${\mu}/{q_{\oplus}^2}$ in the denominator of
$\Lambda_n$ is the size of the heliocentric acceleration of the Earth,
$\simeq 0.6 \; cm\,s^{-2}$.  Thus $|\Lambda_n|$ can be $>1$, and the
coefficient $1-\Lambda_n$ very different from $1$; it may even be
negative. This leads to the conclusion that without taking into
account the topocentric correction the classical method of Laplace is
not a good approximation in the general case\footnote{When
observations from different nights are taken by the same station at
the same sidereal time the topocentric correction in acceleration
cancels out.  In this case the classical Laplace method is a good
approximation.}.

The common procedure when using Laplace's method is to apply a
negative topocentric correction to go back to the geocentric
observation case. However, in doing this some value of $\rho$ is
assumed as a first approximation. If this value is approximately
correct, by iterating the cycle (topocentric correction - Laplace's
determination of $\rho$) convergence is achieved. If the starting
value is really wrong the procedure may well diverge. (\eg a value
$\rho=1\, AU$ is assumed by default and the object is actually
undergoing a close approach to the Earth). Moreover, in this way the
information contained in the parallax is not exploited in an optimal
way. Unless a better way is found to account for the topocentric
correction there are reliability problems discouraging the use of
Laplace's method when processing a large dataset, containing
discoveries of objects of different orbital classes and therefore
spanning a wide range of distances.
 
% ============================================================================

\subsection{Gauss-Laplace equivalence, Topocentric}
% =============================================================================

When taking into account the displacement $\mathbf{P}$ the Taylor expansion
of $\q_i(t)$ of eq.~(\ref{serie terra}) is not applicable.  We need to use
\[
\q_i=\q_2+t_{i2}\dot{\q}_2+\frac{t_{i2}^2}{2}\ddot{\q}_2+{\cal O}(\Delta t^3)
\]
where $\q_2(t)$ and its derivatives contain also $\mathbf{P}(t)$.
By using eq.~(\ref{prodotto vettore}) and assuming $t_{21}=t_{32}$,
eq.~(\ref{B}) and (\ref{A}) become
\[
B(\q_1,\q_3)=\frac{\mu\;\eta}{2}\;t_{21}t_{32}t_{31}^2\;
\hat{\nn}_2\cdot\q_2+{\cal O}(\Delta t^6)
\]
\[
A(\q_1,\q_2, \q_3)
=\frac{q_2^3\;\eta}{2}\;t_{21}t_{32}t_{31}^2\;
\hat{\nn}_2\cdot\ddot{\q}_2+{\cal O}(\Delta t^6)\ .
\]
Note that $\dot\q_2$ does not appear in $A$ at this approximation level.
Thus
\[
h_0=-\frac{A}{B}=
-\frac{q_2^3\ \hat{\nn}_2\cdot\ddot{\q}_2+{\cal O}(\Delta t^2)}{\mu\ 
\hat{\nn}_2\cdot\q_2+{\cal O}(\Delta t^2)}
\]
and once again neglecting $P/q_\oplus$ terms
\[=-\frac{q_2^3\ \hat{\nn}_2\cdot\ddot{\q}_{\oplus 2}}{\mu\ 
\hat{\nn}_2\cdot\q_2}-\frac{q_2^3\ \hat{\nn}_2\cdot
\ddot{\mathbf{P}}_2}{\mu\ \hat{\nn}_2\cdot\q_2}+O(\Delta t^2)=
\]
\[=\frac{q_2^3}{q_{\oplus 2}^3}-\frac{q_2^3\ 
\hat{\nn}_2\cdot\ddot{\mathbf{P}}_2}{\mu\ \hat{\nn}_2\cdot\q_2}
+{\cal O}(\Delta t^2).
\]
Finally
\[
\hat{\nn}_2\cdot\q_2=
q_2\ \hat{\nn}_2\cdot\left(\frac{\q_{\oplus 2}}{q_2}+
\frac{\mathbf{P}_2}{q_2}\right)=
q_2\left(\hat{\nn}_2\cdot\hat{\q}_{\oplus 2}+
{\cal O}\left(\frac{P_2}{q_2}\right)\right)
\]
then
\[
h_0=1-\frac{q_{\oplus 2}^3\ \hat{\nn}_2\cdot
\ddot{\mathbf{P}}_2}{\mu\ \hat{\nn}_2\cdot\q_2}
+{\cal O}(\Delta t^2)
+{\cal O}\left(\frac{P_2}{q_2}\right)=
1-\Lambda_{n2}+{\cal O}(\Delta t^2)+{\cal O}\left(\frac{P_2}{q_2}\right)
\]
where $\Lambda_{n2}$ is the same quantity as $\Lambda_n$ given by
eq.~(\ref{chi}) and computed at $t=t_2$.

The conclusion is that Gauss' method used with the
heliocentric positions of the observer $\q_i$ is equivalent to
Laplace's method with topocentric correction to lowest order in
$\Delta t$ and neglecting the very small term ${\cal O}({P_2}/{q_2})$.
%=============================================================================

\subsection{Problems in Topocentric Laplace's Method}

The results obtained in this Section can be summarized as follows:
contrary to common belief, Gauss' method is not equivalent to
Laplace's unless Gauss' is artificially spoiled by not using the
observer position in eq.~(\ref{B}) and (\ref{A}). In its rigorous
form, Gauss' method accounts for the topocentric correction with a
consistent approximation.
The question arises whether we could account for the topocentric
correction in Laplace's method (without iterations) by adding the term
$\Lambda_n$ from eq.~(\ref{chi}).  Surprisingly, the answer is already
contained in the literature in a 100 year old paper by a famous author
\cite[pag. 177--178]{poincare}.  To summarize the argument of
Poincar\'e, plots showing the shape of the topocentric corrections as
a function of time and a short citation are enough.

\begin{figure}[h]
\centerline{\epsfig{figure=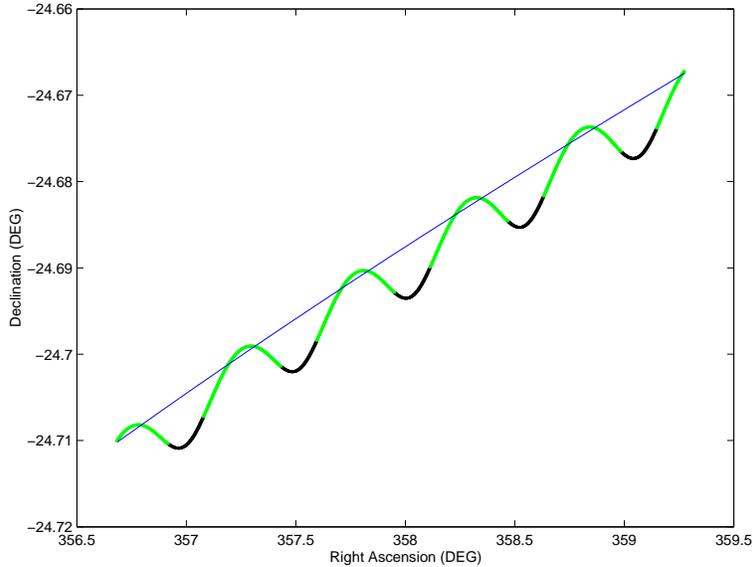,width=10cm}}
\caption{The path in the sky of an approaching NEO. This example is
  (101955) 1999 RQ$_{36}$ as it would have been seen in July 2005 if
  an observatory on Mauna Kea had been observing
  continously. The curve resembling a parabola gives
  simulated observations from the geocenter.}
  \label{fig:path_maunakea}
\end{figure}

Figure~\ref{fig:path_maunakea} shows the simulated path of an
approaching NEO as seen from an observing station (in this example in
Hawaii).  The darker portions of the curve indicate possible
observations that have an altitude $>15^\circ$.  The overall apparent
motion of the asteroid from night to night cannot be approximated
using parabolic motion segments fitted to a single night\footnote{Our
translation of Poincar\'e: \emph{It is necessary to avoid computing
these quantities by starting from the law of rotation of the
Earth.}}. For the geocentric path the parabolic approximation to
$\hat{\brho}(t)$, used by Laplace, would be applicable.

\begin{figure}[h]
\centerline{\epsfig{figure=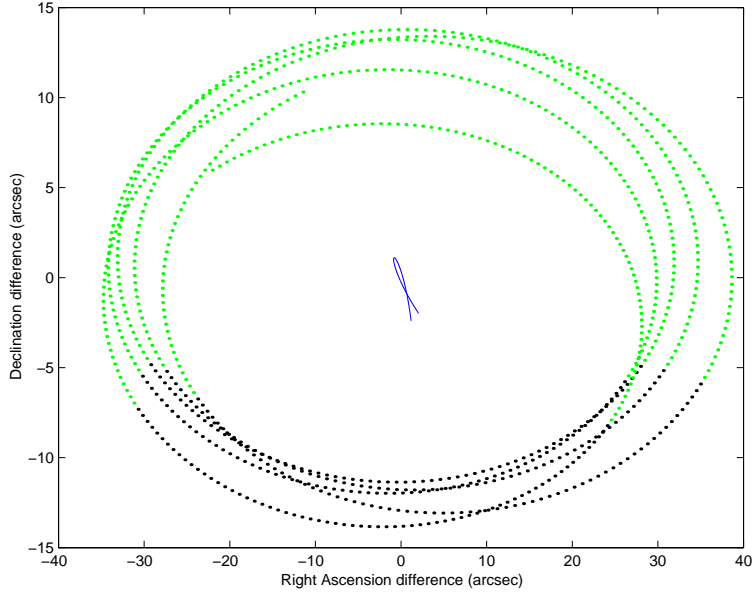,width=10cm}}
\caption{The same data as in the previous figure after removing the
best fitting linear functions of time in both coordinates. In this
case the curves represent the content of information beyond the
attributable. The larger loop is from Mauna Kea while the small curl
near $(0,0)$ is for a geocentric observer. Coordinates are differences
in RA and DEC in radians.}
\label{fig:resid_maunakea}
\end{figure}

Figure~\ref{fig:resid_maunakea} shows graphically that topocentric
observations contain information beyond what is contained in the
average angles and proper motion (the attributable, see
Section~\ref{s:distant}).  Thus, to reduce the observations to the
geocenter by removing the topocentric correction is not a good
strategy.

Poincar\'e suggests computing what we call $\Lambda_n$ by using a value of
$\ddot\mathbf{P}$ obtained by interpolating the values
$\mathbf{P}(t_i)$ at the times $t_i$ of the observations which are not
limited to 3 (one of the advantages of Laplace's method). We have
implemented Poincar\'e's suggestion to improve Laplace's method in our
software system {\sl OrbFit}\footnote{In Version 3.4.2 and later; see
{\tt http://newton.dm.unipi.it/orbfit/}} but we still need to test it
properly.

When the observations are performed from an artificial satellite (such
as the Space Telescope or, in the future, from Gaia) the acceleration
$\ddot{\mathbf{P}}\simeq 900\; cm\;s^{-2}$ and the $\Lambda_n$ and
$\Lambda_v$ coefficients can be up to $\simeq 1,500$. A few hours of
observations extending to several orbits can produce multiple kinks
as in \cite[Figure 1]{marchi} containing important orbital
information.

% =========================================================================
\section{Qualitative Theory, Topocentric}
\label{s:qualitative}

In rectangular heliocentric coordinates $(x,y)$ where
the $x$ axis is along $\hat\q_2$ (from the Sun to the observer) we
have $\rho_2 = \sqrt{q_2^2 + x^2 + y^2 - 2xq_2}$ and $r_2 =
\sqrt{x^2+y^2}$, thus we can consider the function
\[
C_0(x,y) = \frac{q_2}{ \sqrt{q_2^2 + x^2 + y^2 - 2xq_2}}\left[h_0 -
\frac{q_2^3}{(x^2+y^2)^{3/2}}\right]\ .
\]
The dynamical equation eq.~(\ref{dinamicagauss}) can be seen as
describing the level lines $C_0=const$ in a bipolar coordinate system
$(r_2,\rho_2)$.  Note that $C_0=0$ is the \emph{zero circle}
$r=r_0=q/\sqrt[3]{h_0}$ for $h_0>0$ and is otherwise empty.
%If $h_0\neq 1$, this function has a pole of order 1 in $(x,y)=(q_2,0)$ and a
%pole of order 3 in $(x,y)=(0,0)$.
%\begin{equation}
%\left\{
%\begin{array}{ll}
%\displaystyle\frac{\partial C_0}{\partial x}
%  &=\displaystyle\frac{q_2}{\rho_2^3 q_2^5}\left\{\left[3q_2^3\rho_2^2 -
%  (h_0r_2^3 - q_2^3)r_2^2 \right] x + q_2r_2^2(h_0r2^3 - q_2^3)\right\}\cr
%
%&\cr
%
%\displaystyle\frac{\partial C_0}{\partial y}
%  &=\displaystyle\frac{q_2}{\rho_2^3 q_2^5}\left[3q_2^3\rho_2^2 - (h_0r_2^3 -
%  q_2^3)r_2^2 \right] y \cr
%\end{array}
%\right.
%\end{equation}
A simple computation of the partial derivatives of $C_0$ shows that
the only stationary points of $C_0$ are the pairs $(x,y)$ with $y=0$
and $x$ a solution of the equation $ (h_0|x|^3 - q_2^3)x =
3q_2^3(x-q_2) $.  For $h_0\leq 0$ it has only one solution, $x_1$,
with $0< x_1 < q_2$.  For $h_0>0$ there is always at least one
solution $\bar x <-r_0 < 0$.  If $0< h_0< 1$ there are two additional
solutions, $x_1$ and $x_2$ such that $0<x_1 < q_2< r_0 < x_2$.  For
$h_0>1$ there are no positive solutions\footnote{The quantity $C$
appearing in the topocentric Laplace's method defines exactly the same
function of $(x,y)$, with $h_0$ replaced by $1-\Lambda_n$.}.
%\clearpage
\begin{figure}[t]
\centerline{\epsfig{figure=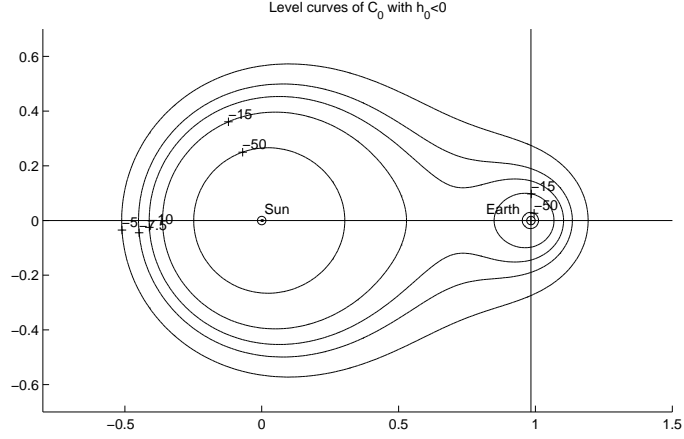,width=9cm}}
\caption{Level curves of $C_0(x,y)$ for $h_0=-0.5$. Note there is no
zero circle.}
\label{fig:h0lt0}
\end{figure}

The function $C_0(x,y)$ has a pole of order 3 at
$(x,y)=(0,0)$, with $\lim_{r_2\to 0} C_0 = -\infty$, and a pole of
order 1 at $(q_2,0)$ with  $\lim_{r_2\to 0} C_0 =+\infty$
for $h_0>1$ and $=-\infty$ for $h_0<1$. For $h_0=1$
there is at $(q_2,0)$ a more complicated singularity: as shown by
Figure~\ref{fig:charlier2} there is no unique limit value for
$\rho_2\to 0$.

\subsection{Topology of the level curves of $C_0(x,y)$}

\begin{figure}
\centerline{\epsfig{figure=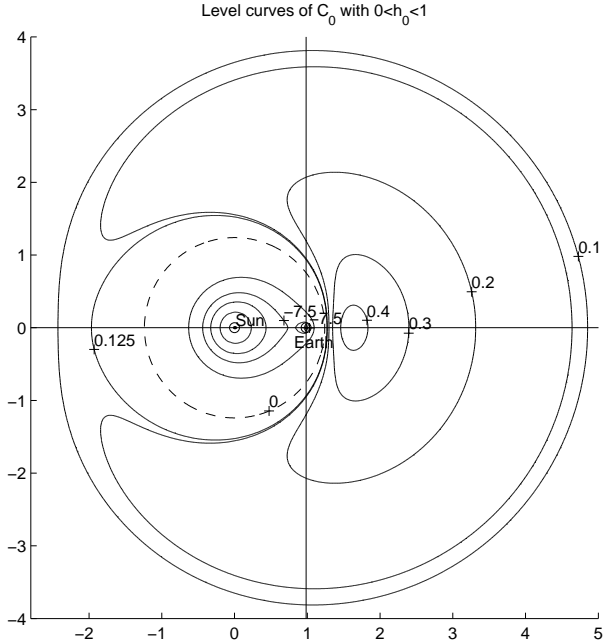,width=8cm}}
\caption{Level curves of $C_0(x,y)$ for $h_0=0.5$, including the zero
circle (dashed).}
\label{fig:h0lt1}
\end{figure}

\begin{figure}[h]
\centerline{\epsfig{figure=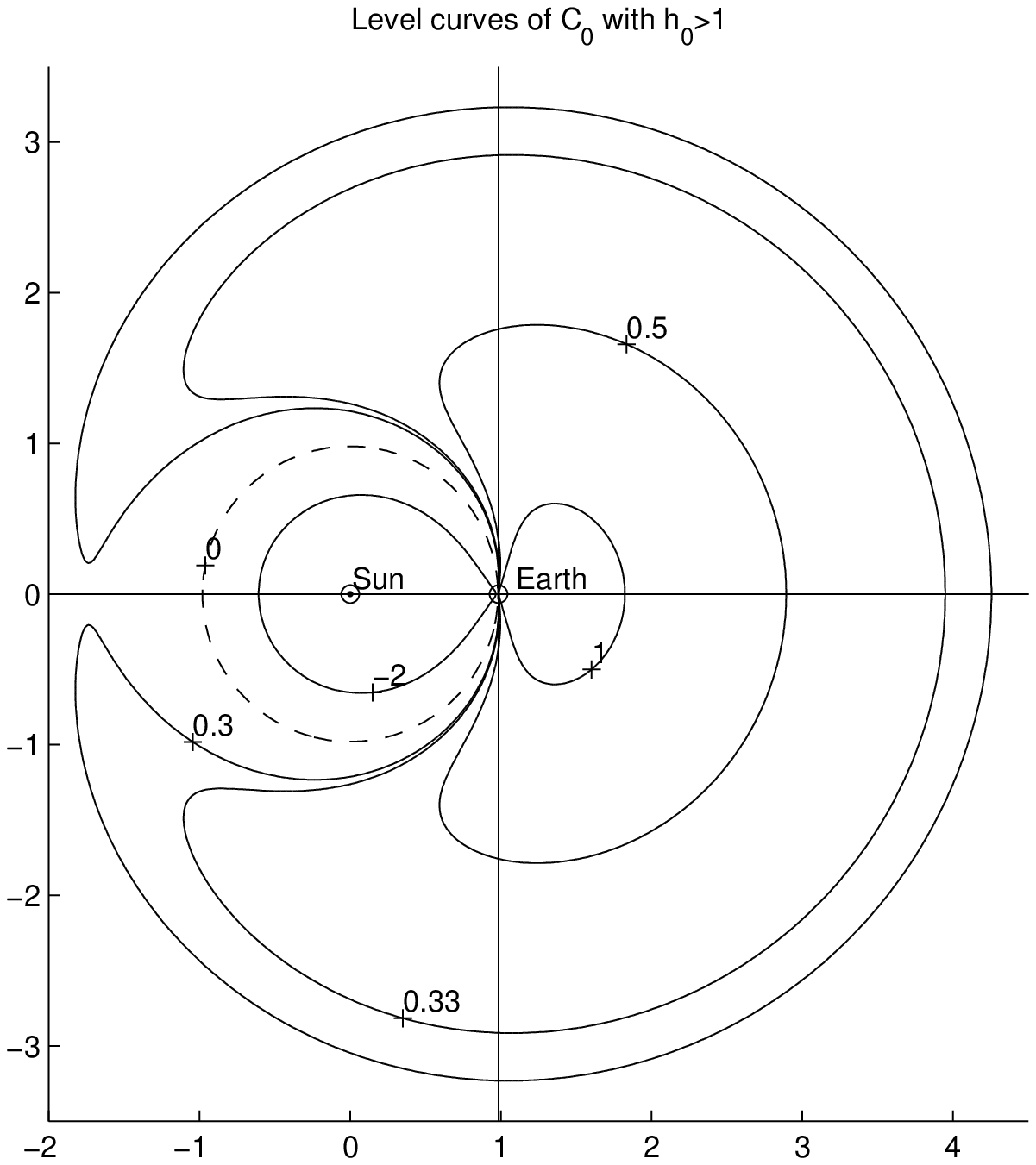,width=9cm}}
\caption{Level curves of $C_0(x,y)$ for $h_0=1.1$, including the zero
circle (dashed).}
\label{fig:h0gt1}
\end{figure}

The qualitative behavior of the level lines of $C_0(x,y)$ is different
in the three cases $h_0\leq 0$ (Figure~\ref{fig:h0lt0}), $0<h_0<1$
(Figure~\ref{fig:h0lt1}) and  $h_0>1$ (Figure~\ref{fig:h0gt1}).

%For $h_0<0$ we have $\lim_{r_2\to 0} C_0 = -\infty$ and
%$\lim_{\rho_2\to 0} C_0 = -\infty$ and $(x_1,0)$ is a saddle as shown in
%Figure~\ref{fig:h0lt0}.

%For $0<h_0<1$ we have $\lim_{r_2\to 0} C_0 = -\infty$ and
%$\lim_{\rho_2\to 0} C_0 = -\infty$. The stationary points are as
%follows: $(\overline x,0)$ is a saddle, $(x_1,0)$ is also a saddle 
%and $(x_2,0)$ corresponds to the absolute maximum value of $C_0$
%(see Figure~\ref{fig:h0lt1}).

%For $h_0>1$ we have $\lim_{r_2\to 0} C_0 = -\infty$ and
%$\lim_{\rho_2\to 0} C_0 = +\infty$. The only critical point is
%$(\overline x, 0)$ which is a saddle and the only possible
%qualitative behavior is shown in Figure~\ref{fig:h0gt1}.

%We need to use the \emph{singular curve} which is obtained from the
%geometric and dynamic equations of Gauss' method with an argument
%perfectly analogous to eq.~(\ref{sing1}) and (\ref{sing2}).  In
%heliocentric polar coordinates, $(r_2,\phi_2)$, the singular curve has
%the form
%\begin{equation}
%h_0 r_2^3 = q_2^3\left(4 - 3\frac{q_2}{r_2}\cos\phi_2\right)\ .
%\label{singc_topcorr}
%\end{equation}
The number of solutions of the dynamical equation (\ie along a
fixed topocentric direction) can be understood by evaluating the degree
8 polynomial (\ref{polinomio gauss}) on the zero circle
\[
P(r_0)=C_0^2\; \frac{q^8}{h_0^{8/3}}\; \left(1-h_0^{2/3}\right)\ .
\]
%and by considering the sign of $C_0$. 
 \begin{table}[h]
   \caption{The number of preliminary orbit solutions, the
   columns give: [1] the number of preliminary orbit solutions (for a
   given $C$ and $\epsilon$), [2] the number of positive roots of the
   polynomial equation (\ref{polinomio gauss}), [3] the number of
   spurious roots.}
   \begin{center}  
    \begin{tabular}{lcccc}
      & &{[1]} &{[2]} &{[3]}\\
      \hline
      $h_0\le 0$ &${\cal C}<0$ &1 or 3 &1 or 3 &0 \\
      &${\cal C}>0$ &0 &1 or 3 &1 or 3 \\
      \hline
      $0<h_0<1$ &${\cal C}<0$ &1 or 3 &1 or 3 &0 or 2 \\
      &${\cal C}>0$ &0 or 2 &1 or  3 &1 or 3 \\
     \hline
      $h_0=1$ &${\cal C}<0$ &0 or 1 or 2 &1 or 3 &1 or 2 or 3\\
              &${\cal C}>0$ &0 or 1 or 2 &1 or 3 &1 or 2 or 3\\
     \hline
      $h_0>1$ &${\cal C}<0$ &0 or 2 &1 or 3 &1 or 3 \\
      &${\cal C}>0$ &1 or 3 &1 or 3 &0 or 2 \\
    \end{tabular}
    \end{center}
  \label{t:nosol}
  \end{table}

We summarize the possible numbers of solutions, for a given direction
of observation $\epsilon$, in the different cases, depending upon the
value of $h_0$ and the sign of $C$, in Table~\ref{t:nosol}.  Note that
in the Table we are not making the assumption of Charlier, that some
solutions must exist, for the reasons given in
Section~\ref{s:charlier}. By spurious we mean a root of the polynomial
(\ref{polinomio gauss}) corresponding to $\rho\leq 0$ in
eq. (\ref{dinamicagauss}). This is not a complete qualitative theory
replacing Charlier's for $h_0=1$, but already shows that the number of
solutions can be quite different from the classical case \eg, 2
solutions near opposition and up to 3 at low elongation. For a fully
generalized qualitative theory see \cite{gronchistmalo}.

%\clearpage

%\clearpage

%For $h_0<0$ we have $r_0<0$ (no zero circle).  With $P(0)<0$ and
%$\lim_{r\to+\infty}P(r)=+\infty$ it implies that there can be either 1 or
%3 positive solutions (only on the singular curve are there 2
%solutions, one with multiplicity two). From Figure~\ref{fig:h0lt0} it
%is clear that the 3 roots can occur only at low solar elongations
%while one root is found near the opposition direction.

%For $0<h_0<1$ we have $P(0)<0,\; P(r_0)>0$ and there is always at
%least a positive root $r_1<r_0$. For $C_0>0$ the root $r_1$ is
%spurious.  Thus, there can be either 0 or 2 acceptable solutions (1 on
%the singular curve). From Figure~\ref{fig:h0lt1} we can see that it is
%possible to have 2 solutions even near opposition for appropriate
%values of $C_0>0$. For $C_0<0$ we can have either 1 or 3 solutions (2
%on the singular curve).

%For $h_0>1$ we have $P(0)<0, P(r_0)<0$ and there is always a root
%$r_1>r_0$.  For $C_0>0$ this root is not spurious and the number of
%acceptable roots can be either 1 or 3 (2 on the singular curve).  For
%$C_0<0$ $r_1$ is spurious and there can be either 0 or 2 acceptable
%solutions (1 on the singular curve).

%The discussion above is not a complete qualitative theory replacing
%the one of Charlier for $h_0=1$ but already shows that the number of
%solutions can be quite different from the classical case \eg, 2
%solutions near opposition and up to 3 at low elongation. For a full
%generalized Charlier theory see \cite{gronchistmalo}.

\subsection{Examples}
\label{s:examples}

\begin{figure}[h]
\centerline{\epsfig{figure=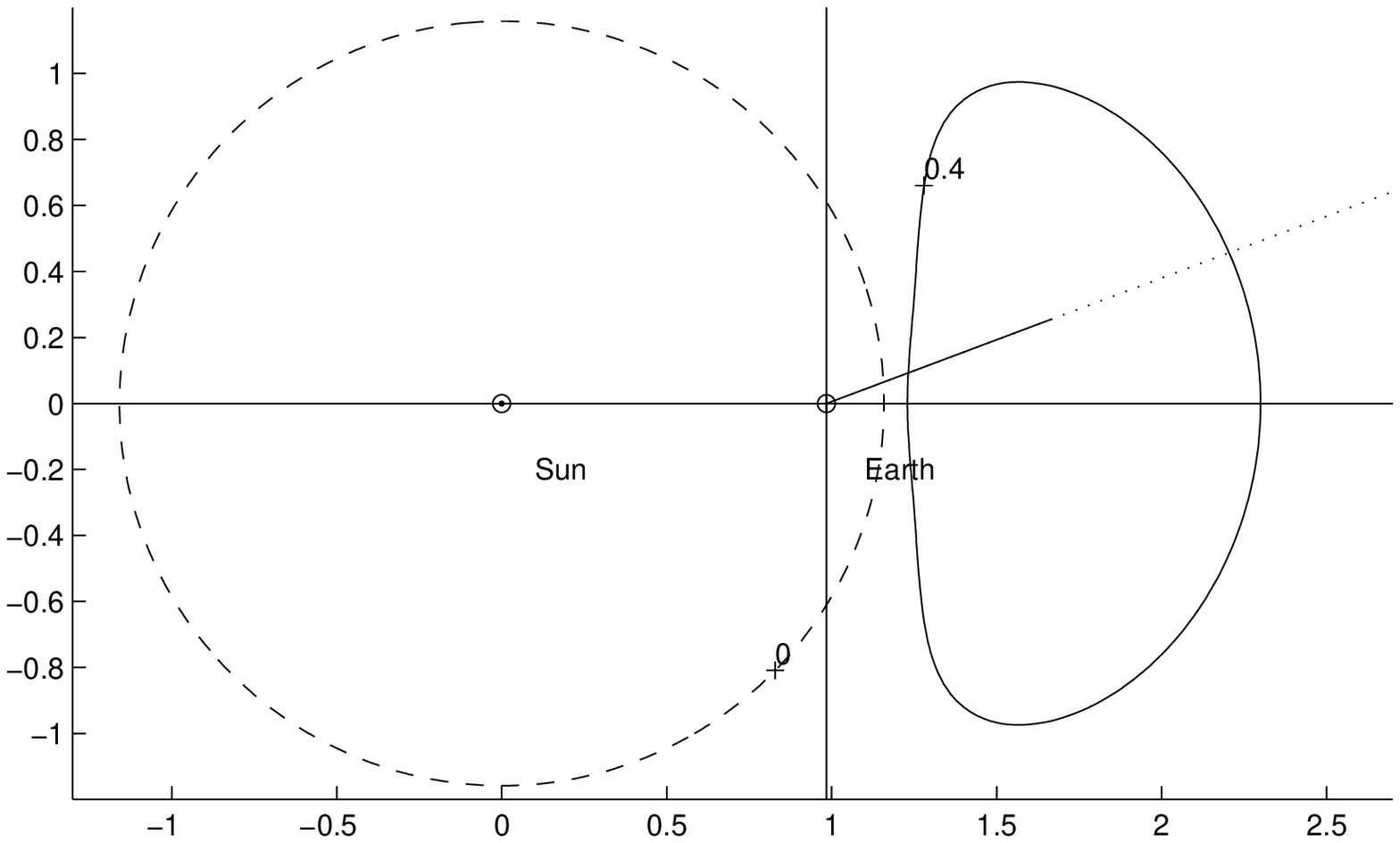,width=10cm}}
\caption{An example with two solutions near opposition: for
$h_0=0.613$ the direction of observation (dotted) has two
intersections with the level curve $C_0(x,y)=0.4$ (continuous); the
zero circle is dashed.  The positions in the observation
direction with a bounded orbit are drawn as a continuous
line.}
\label{fig:2solopp}
\end{figure}

We would like to find examples in which the additional solutions with
respect to the classical theory by Charlier are useful.  That is,
cases in which the additional solutions provide a preliminary orbit
closer to the true orbit and therefore more suitable as a first guess
for the differential corrections procedure.

An example in which there are two solutions while observing in a
direction close to the opposition is shown in
Figure~\ref{fig:2solopp}.  Of the two intersections of the observing
half line with the relevant level curve of $C_0$, the one leading to a
useful preliminary orbit is the nearer one which has $\rho_2=0$ as
counterpart in the classical theory. The farther one leads to a
preliminary orbit with $e\simeq 10$. We have used the formulae of
\cite{ons1} to compute the maximum possible $\rho_2$ along the
observation direction compatible with $e\leq 1$.

Another interesting feature of this example is that the preliminary
orbit using the nearer solution has residuals of the 6 observations
with RMS $=66$ arcsec while the one using the farther solution has RMS
$=2.5$ arcsec. This implies that if only one preliminary solution
were passed to the next processing step by selecting the one with
lowest RMS the good solution would be discarded.

\begin{figure}[h]
\centerline{\epsfig{figure=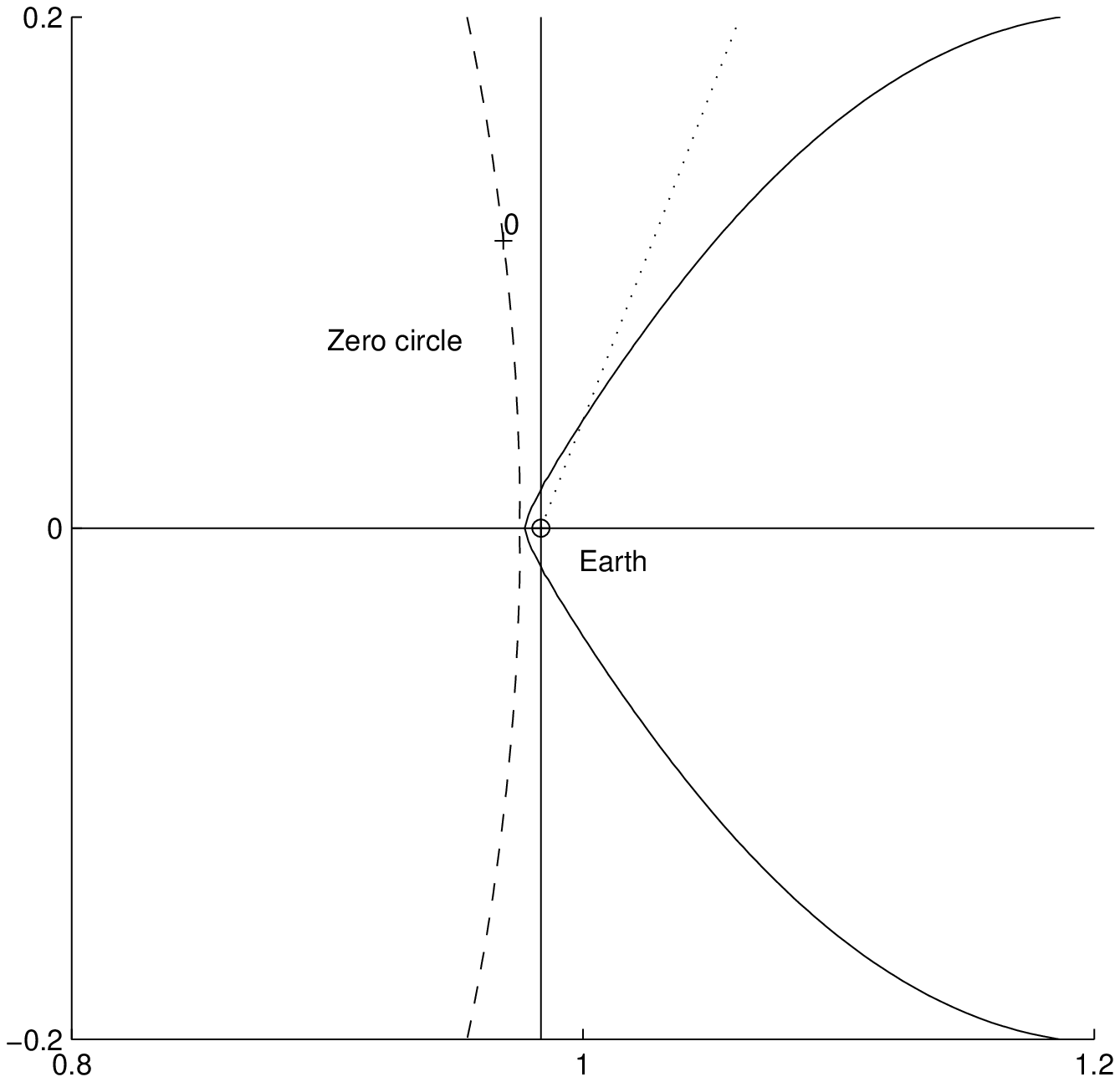,width=6.5cm}
\epsfig{figure=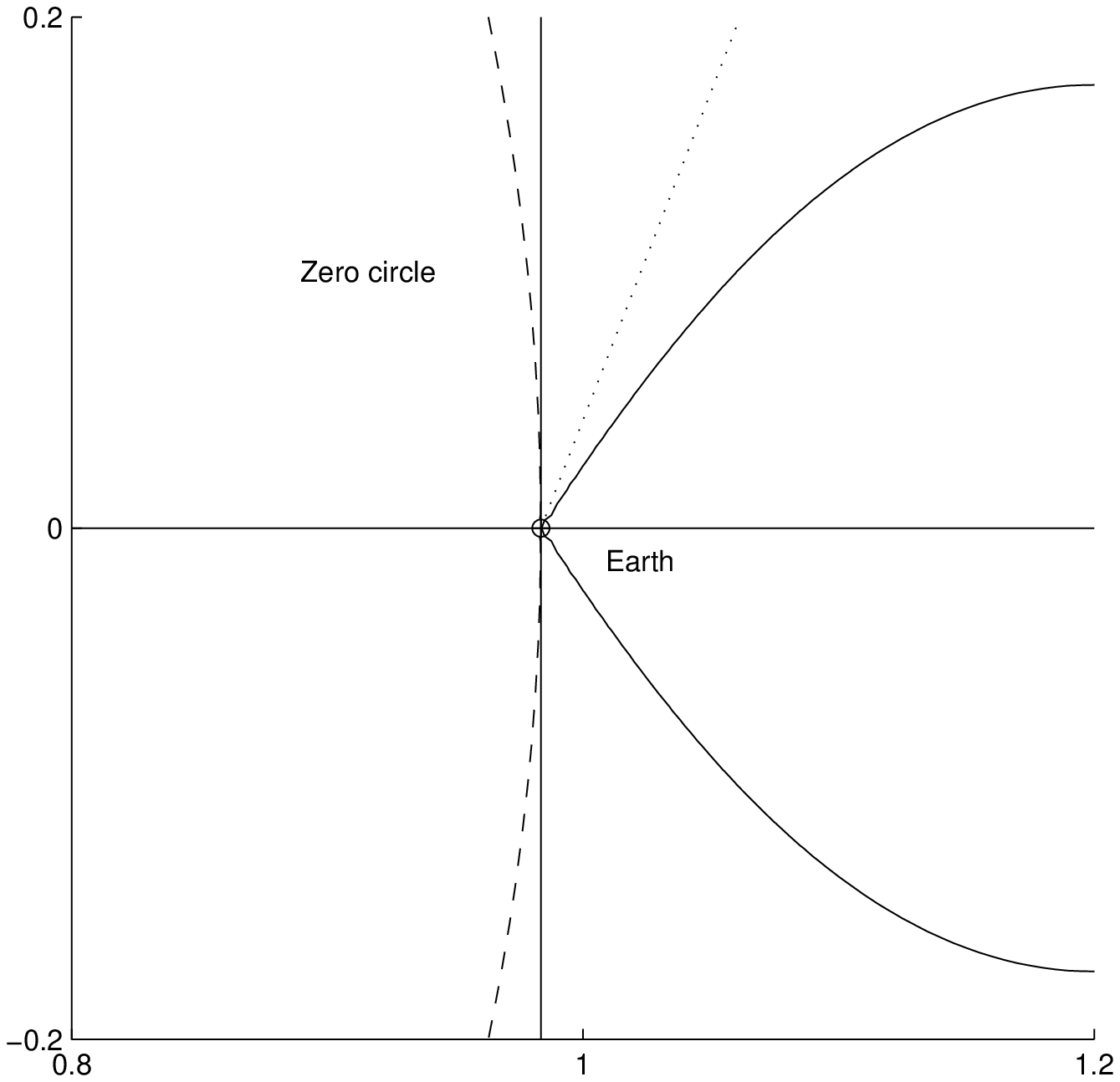,width=6.5cm}}
\caption{For the preliminary orbit of 2002 AA$_{29}$ the relevant
level curve ($C_0=1.653$) is shown (continuous) in the same plane of
Figure~\ref{fig:charlier2}; the zero circle (dashed) and the
observation direction (dotted) are also shown. Left: using the actual
value $h_0=1.025$.  Right: using a value of $h_0=1$ that does not
account for the topocentric correction.}
\label{fig:aa29}
\end{figure}

To find a significant example with 3 solutions is not easy because in
many cases the third solution, the nearest to the observer, has
$\rho_2$ too small for the heliocentric 2-body approximation to be
applicable. A value $\rho_2\leq 0.01$ AU corresponds to the
\emph{sphere of influence} of the Earth, i.e., the region where the
``perturbation'' from the Earth is actually more important that the
attraction from the Sun. Thus, a solution with such a small $\rho_2$
must be discarded, because the approximation used in Gauss' and
Laplace's method is not valid.

To show how our arguments on the number of solutions applies to a real
case (as opposed to a simulation as in the example above) we have
selected the asteroid 2002 AA$_{29}$ and used observations from the
first three nights (9, 11 and 12 January 2002). With the values
$C_0=1.653\; ,\; h_0=1.025$ we obtain from the observations and an
elongation $\simeq 111^\circ$ there is only one solution with
$\rho_2=0.0045$ (see Figure~\ref{fig:aa29}, left), which easily leads
to a full least squares solution with $\rho_2=0.044$.  Although the
value of $h_0$ is not very far from 1 the existence of the solution
depends critically on $h_0-1\neq 0$. If the value of $h_0$ had been
set to 1 we would find no solution (see Figure~\ref{fig:aa29}, right).

% ============================================================
\subsection{Implementation issues}
\label{s:implementation}

We need to implement the algorithms discussed in this paper for the
computation of preliminary orbits in a way which is suitable for a
large observation data set; we need to satisfy three requirements.

The first requirement is to obtain the solutions to the polynomial
equations such as (\ref{polinomio gauss}) in a way which is
fast and reliable in providing the number of distinct
real solutions. In this way we can fully exploit the understanding on
the number of solutions (with topocentric correction) which we have
achieved in this section. This is made possible by the algorithms
computing the set of roots of a polynomial equation at once (as a
complex vector) and with rigorous upper bounds for the errors
including the ones generated by roundoff.  We use the algorithm by
\cite{bini} and the corresponding public domain software\footnote{For
the Fortran 77 version {\tt http://www.netlib.org/numeralgo/na10}. For
Fortran 90 {\tt http://users.bigpond.net.au/amiller/pzeros.f90 .}}.

The second requirement is to improve the preliminary orbit as obtained
from the solutions of the degree 8 polynomial equations in such a way
that it is as close as possible to the ``true'' solution to be later
obtained by differential corrections. There is such an immense
literature on this topic that in this paper it is not even appropriate
to discuss the references. Conceptually, as shown by \cite{celletti},
each step in the iterative procedures used in differential corrections
can be shown to increase the order in $\Delta t$ of the approximation
to the exact solutions of the 2-body equations of motion. However,
\cite{celletti2} have also shown that an iterative \emph{Gauss map}
can diverge when the solution of the degree 8 equation is far from the
fixed point of the iterative procedure, outside of its convergence
domain.

We have implemented one of the available iterative improvement
algorithms for Gauss' method and have found that it
provides in most cases a preliminary orbit much closer to the least
squares solution and therefore a more reliable first guess for the
least squares algorithms.
%\footnote{It also increases speed because the
%iterative Gauss map is much faster than even a single extra iteration of
%differential corrections.}. 
We have also found that the Gauss map diverges in a small fraction of
the test cases but still often enough to significantly decrease the
efficiency\footnote{See Section~\ref{s:tests} for the definition of
the metric.} of the algorithm. In some cases the number of
orbits for which the Gauss map converges is less than the number of
solutions of the degree 8 equations. It can happen that one of the
lost degree 8 solutions was the one closest to the ``true'' orbit and
the only one which can be used to obtain the best least squares
solutions.  One method to obtain the highest efficiency without an
inordinate increase in the computational cost is to run two
iterations, one with and one without the Gauss map.

The third requirement is to use modified differential corrections
iterative algorithms with larger convergence domains in such a way
that even when the geodetic curvature (contained in the coefficients
$C$ and $C_0$ of the two methods) is poorly constrained by the
available observations (because the arc length on the celestial sphere
is too short) the very rough preliminary orbit solution can lead to
a least squares solution. This is discussed in the next section.

\section{Weak preliminary orbits}
\label{s:distant}

An essential difference between the classical works on preliminary
orbits and the modern approach to the same problem is that the effects
of the astrometric errors cannot be neglected, especially in the
operating condition of modern surveys: they use shorter observed arcs,
thus the deviations of the observed path on the celestial sphere from
a great circle may not be significant.

\subsection{Uncertainty of Curvature}
\label{s:unccurv}

The explicit computation of the two components of curvature of
interest for orbit determination, geodesic curvature $\kappa$ and
along track acceleration $\dot\eta$, can be performed by using the
properties of the orthonormal frame (\ref{movingframe}) by
straightforward computation using the Riemannian structure of the
unit sphere \cite[Section 6.4]{discovery}. The results are
\begin{eqnarray}
\kappa&=&\frac{1}{\eta^3}\;\left\{
(\ddot\delta\;\dot\alpha-\ddot\alpha\;\dot\delta)\;\cos\delta
+\dot\alpha\;\left[\eta^2+(\dot\delta)^2\right]\;\sin\delta\right\}
=\kappa(\alpha,\delta,\dot\alpha,\dot\delta,\ddot\alpha,\ddot\delta)\ \ \ 
\label{geocurv}\\
\dot\eta&=& \frac 1\eta \;\left[
\ddot\alpha\;\dot\alpha\;\cos^2\delta+\ddot\delta\;\dot\delta
-(\dot\alpha)^2\;\dot\delta\;\cos\delta\;\sin\delta\right]
=\dot\eta(\alpha,\delta,\dot\alpha,\dot\delta,\ddot\alpha,\ddot\delta)\ .
\label{accel}
\end{eqnarray}
Given these explicit formulae it is possible to compute the covariance matrix 
of the quantities $(\kappa, \dot\eta)$ by propagation of the covariance matrix
of the angles and their derivatives with  the matrix of partial derivatives
for $\kappa$ and $\dot\eta$
\begin{equation}
\Gamma_{\kappa, \dot\eta}=
\derparz{(\kappa,\dot\eta)}{(\alpha,\delta,\dot\alpha,\dot
\delta,\ddot\alpha,\ddot\delta)} \; 
%\matrdue{\Gamma_\alpha}{\underline 0}{\underline 0}{\Gamma_\delta} \;
\Gamma_{\alpha,\delta}\;
\left[\derparz{(\kappa,\dot\eta)}{(\alpha,\delta,\dot\alpha,\dot
\delta,\ddot\alpha,\ddot\delta)}\right]^T\ .
\label{gammaked}
\end{equation}
The covariance matrix $\Gamma_{\alpha,\delta}$ for the angles and
their first and second derivatives is obtained by the procedure of
least squares fit of the individual observations to a quadratic
function of time. The partials of $\kappa$ and $\dot\eta$ are given
below (note that the partials with respect to $\alpha$ are zero).
\begin{eqnarray*}
\frac{\partial \kappa}{\partial \delta}&=& 
-\frac{1}{\eta^5}\Bigl[ -2\,{{\it
	\dot{\alpha}}}^{3}\cos^2\delta\;\sin\delta\; {\it
      \ddot{\delta}}+\sin\delta\;\ddot{\delta}\,
      \dot{\alpha}\,{\dot{\delta}}^{2}+2\,{\dot{\alpha}}^{2}
      \cos^2\delta\;\sin\delta\;{\ddot{\alpha}}\,\dot{\delta}- 
\\
&-& \sin\delta\;{\it \ddot{\alpha}}\,{{\it
	\dot{\delta}}}^{3}-{{\it \dot{\alpha}}}^{5}\cos^3\delta\;
    -4\,{{\it \dot{\alpha}}}^{3}\cos\delta\;{{\it
	\dot{\delta}}}^{2}+{{\it \dot{\alpha}}}^{ 3}\cos^3\delta\;
{\dot{\delta}}^{2}-2\,{\it \dot{\alpha}}\,\cos\delta\;{\dot{\delta}}^{4}\Bigr]
\end{eqnarray*}
\[
\frac{\partial \dot{\eta}}{\partial \delta}=
 -\frac{\dot{\alpha}}{2\;\eta^3} \Bigl[\sin(2\delta)\,\left(
   {\dot{\alpha}}^{2}\;\ddot{\alpha}\;\cos^2\delta
   +2\;{\dot{\delta}}^{2}\;\ddot{\alpha}
   -\dot{\alpha}\;\dot{\delta}\;\ddot{\delta}
\right)
      +2\,\dot{\alpha}\;{\dot{\delta}}^{3}\,\cos(2\delta)+
      2\,{\dot{\alpha}}^{3}\dot{\delta}\;\cos^4\delta\Bigr]
\]
\[
\frac{\partial \kappa}{\partial \dot{\alpha}}= 
\frac{1}{\eta^5}\left[ 
-\dot\alpha\;\cos^3\delta\, \left(
2\,\dot{\alpha}\;\ddot{\delta}
-3 \,\dot{\delta}\;\ddot{\alpha}
\right)+ {\dot{\delta}}^{2}\,\left(
\ddot{\delta}\,\cos\delta-
{\dot{\alpha}}^{2}\cos^2\delta\;\sin\delta
+2\,{\dot{\delta}}^{2}\;\sin\delta
\right)
\right]
\]
\[
\derparz{\dot{\eta}}{\dot{\alpha}}= 
-\frac {\cos\delta\;\;\dot{\delta}}{ \eta^3 }\;\left[ \,
-\cos\delta\;{\;\ddot{\alpha}}\,
      \dot{\delta} +{\dot{\alpha}}^{3}\sin\delta\;
\cos^2\delta\;+2\,{\dot{\alpha}} \,\sin\delta\;{
	\dot{\delta}}^{2}+\cos\delta\;{\ddot{\delta}}\,
      \dot{\alpha}\right]
\]
\[
\derparz{\kappa}{\dot{\delta}}= -\frac 1{\eta^5 } \left[
\cos\delta\,\left({\dot{\alpha}}^{2}\,\ddot{\alpha}\,\cos^2\delta
-2\,{\dot{\delta}}^{2}\;\ddot\alpha
+3\,\dot{\alpha}\;\dot{\delta}\;\ddot{\delta}\right)
-\dot\alpha\;\dot\delta\;\sin\delta\,
\left({\dot{\alpha}}^{2}\,\cos^2\delta-2\,{\dot{\delta}}^{2}
\right)\right]
\]
\[
\derparz{\dot{\eta}}{\dot{\delta}}= -
\frac {\dot{\alpha}\,\cos^2\delta} {\eta^3}\;\left[
      -\ddot{\delta}\,\dot{\alpha}+{\dot{\alpha}}^{3}\cos\delta\;\sin\delta\;
+\ddot{\alpha}\,\dot{\delta} \right]
\]
%%%%%%%%%%%%%%second derivatives
\[
\derparz{\kappa}{\ddot{\alpha}}= -\frac {\dot{\delta}\,\cos\delta\;}{\eta^3} 
\eqspa
\frac{\partial \kappa}{\partial \ddot{\delta}}= 
\frac {\dot{\alpha}\,\cos\delta\;}{ \eta^3 }
\]
\[
\frac{\partial \dot{\eta}}{\partial \ddot{\alpha}}= {\frac {{\it
\dot{\alpha}}\,\cos^2\delta\;}{ \eta }}
\eqspa
\frac{\partial \dot{\eta}}{\partial \ddot{\delta}}= {\frac {{\it
\dot{\delta}}}{ \eta }} \ .
\]
Note that the last four of the partials above, the $2\times 2$
matrix $\partial (\kappa,\dot\eta)/\partial(\ddot\alpha,
\ddot\delta)$, contribute to the principal part of the covariance of
$(\kappa, \dot\eta)$ for short arcs as discussed in the next
subsection.

We use a full computation of the covariance matrix without
approximations to assess the significance of curvature by using the
formula from \cite{discovery}
\begin{equation}
\chi^2=\vecdue{\kappa}{\dot\eta}^T\;\Gamma_{\kappa,\dot\eta}^{-1}\;
\vecdue{\kappa}{\dot\eta}
\label{signcurv}
\end{equation}
and we assume that the curvature is \emph{significant} if
$\chi^2>\chi^2_{min}=9$.  

\subsection{The Infinite Distance Limit}

The problem of low values of $C$ can occur in two ways: near the zero
circle and for large values of both $\rho$ and $r$. On the other hand,
the uncertainty in the estimates of the deviations from a great circle
will depend upon the length of the observed arc (both in time
$\Delta t$ and in arc length $\sim \eta\,\Delta t$). For short
observed arcs it may be the case that the curvature is not
significant.  Then the preliminary orbit algorithms will yield
inaccurate preliminary orbits which may fail as starting guesses for
differential corrections.

We will now focus on the case of distant objects.  We would like to
estimate the order of magnitude of the uncertainty in the computed
orbit with respect to the small parameters $\nu, \tau, b$ where $\nu$
is the astrometric accuracy of the individual observations (in
radians) and $\tau= n_\oplus \Delta t$, $b=\frac{q_\oplus}{\rho} $ are
small for short observed arcs and for distant objects, respectively.
Note that the proper motion $\eta$ for $b\to 0$ has principal part
$n_\oplus\, b$ -- the effect of the motion of the Earth. The
uncertainty in the angles $(\alpha, \delta)$ and their derivatives can
be estimated as follows
\[
\Gamma_{\alpha,\delta}={\cal O}(\nu)\eqspa
\Gamma_{\dot\alpha,\dot\delta}={\cal O}(\nu \tau^{-1})\eqspa
\Gamma_{\ddot\alpha,\ddot\delta}={\cal O}(\nu \tau^{-2})\ .
\]
The uncertainty of the curvature components $(\kappa,\dot\eta)$ should
be estimated by the propagation formula (\ref{gammaked}) but it can
be shown that the uncertainty of $(\delta,\dot\alpha,\dot\delta)$
contributes with lower order terms. Thus we use the estimates
\[
\derparz{(\kappa,\dot\eta)}{(\ddot\alpha,\ddot\delta)}=
\matrdue{{\cal O}(b^{-2})\;n_\oplus^{-2}}{{\cal O}(b^{-2})\;n_\oplus^{-2}}
        {{\cal O}(1)}{{\cal O}(1)}
\]
and obtain
\[
\Gamma_{\kappa,\dot\eta}=\nu\;
\matrdue{{\cal O}(b^{-4}\tau^{-2})}{{\cal O}(b^{-2}\tau^{-2})n_\oplus^2}
{{\cal O}(b^{-2}\tau^{-2})n_\oplus^2}{{\cal O}(\tau^{-2})n_\oplus^4}\ .
\]
%As an example, for $\nu=10^{-6}$ and an observed arc of $\Delta t=10$
%days for an object at $\rho=40$ AU, $\sqrt{\Gamma_{\kappa\kappa}}\simeq 54$. 

To propagate the covariance to the variables $(\rho, \dot\rho)$ we use
the equation, obtained by eliminating $r$ between eq.~(\ref{equazione
geometrica}) and (\ref{dinamicagauss}), an implicit equation
connecting $C$ and $\rho$
\begin{equation}\label{implicita1}
F(C,\rho)= C\ \frac{\rho}{q_\oplus}\ +\frac{q_\oplus^3}{(q_\oplus^2 +
\rho^2 + 2q_\oplus\rho\cos\varepsilon)^{3/2}}\ -1+\Lambda_n=0 \ .
\end{equation}
For $b\rightarrow 0$ we have $C\; b^{-1}\rightarrow 1$ and thus
$C\rightarrow 0$ and is of the same order as the small parameter
$b$. Although $C$ depends upon all the variables
$(\alpha,\delta,\dot\alpha,\dot \delta,\ddot\alpha,\ddot\delta)$, its
uncertainty mostly depends upon the uncertainty of $\kappa$ and thus,
ultimately, upon the difficulty in estimating the second derivatives
of the angles.  Next, we compute the dependence of the uncertainty of
$(\rho, \dot\rho)$ upon the uncertainty of $(\kappa, \dot\eta)$.
From the derivatives of the implicit function $\rho(\kappa)$, assuming
$\cos\varepsilon, \eta, \hat\nn$ to be constant and keeping only the
term of lowest order in $q/\rho$,
\[
\frac{\partial \rho}{\partial \kappa}=
-\frac{\eta^2\;q^4}{\mu\;\hat{\q_\oplus}\cdot\hat{\nn}}\
\frac{\rho}{q_\oplus\;C}+{\cal O}\left(\frac{q^3}{\rho^3}\right)=
q_\oplus\;{\cal O}(1)\ .
\]
In the same way from (\ref{chi_v}) we deduce $ \dot\eta=
n_\oplus^2\,{\cal O}(b) $ and the estimates for the partial
derivatives
\[
\derparz{\dot\rho}{\kappa}=
n_\oplus\;q_\oplus\;{\cal O}(1) \eqspa
\derparz{\dot\rho}{\dot\eta}=\frac{q_\oplus}{n_\oplus} \; {\cal O}(b^{-2})\ .
\]
For the covariance matrix
\[
\Gamma_{\rho,\dot\rho}=\derparz{(\rho,\dot\rho)}{(\kappa,\dot\eta)}\;
\Gamma_{\kappa,\dot\eta}\;
\left[\derparz{(\rho,\dot\rho)}{(\kappa,\dot\eta)}\right]^T
\]
we compute the main terms of highest order in $b^{-1}, \tau^{-1}$ as
\begin{equation}
\Gamma_{\rho,\dot{\rho}}=\nu\;b^{-3}\;\tau^{-2}\;
\matrdue{q_\oplus^2\;{\cal O}(1)}{q_\oplus^2\;n_\oplus\;{\cal O}(1)}
{q_\oplus^2\;n_\oplus\;{\cal O}(1)}{q_\oplus^2\;n_\oplus^2\;{\cal O}(1)}\ .
\label{covrrd}
\end{equation}
In conclusion, if the variables $(\rho, \dot\rho)$ are measured in the
appropriate units (AU for $\rho$ and $n_\oplus$ AU for $\dot\rho$)
the uncertainties are of the same order and there is no reason to
suppose that one of the two will be better determined than the other.

This conclusion is different from the one of \cite{bernstein}.
We are making no assumption about the orbit, just on the
distance of the observed object, \eg the proper motion may not be
well aligned with the ecliptic as is the case for a low eccentricity,
low inclination ``classical TNO''. This is due to the fact that our
main concern is reliability.  We do not want to use an
orbit computation method which might preferentially fail on unusual
orbits, \eg for a long period (even hyperbolic) comet discovered
at large distance. We do agree with \cite{bernstein} on the fact that
for a TNO observed only over an arc shorter than one month there is
very often an approximate degeneracy that forces the use of a
constrained orbit (with only 5 free parameters).  We only claim that
the \emph{weak direction}, which is essentially in the
$(\rho,\dot\rho)$ plane, may vary and is generally not along the
$\dot\rho$ axis \cite[Figures 3-6]{ons2}. In Section~\ref{s:smalltest}
we confirm this argument by a numerical test on TNOs.

\subsection{From Preliminary Orbits to Least Square Solutions}
\label{s:pretolsq}

The procedure to compute an orbit given an observed arc with $\geq 3$
nights of data (believed to belong to the same object) begins with the
solution of the degree 8 equation~(\ref{polinomio gauss}) and ends
with the differential corrections iterations to achieve a full least
squares orbit (with 6 solved parameters). However, for algorithms more
efficient than the classical ones there are up to four intermediate
steps.

We need the definition of \emph{Attributable} \cite{ident4}: the set
of 4 variables $(\alpha, \delta, \dot\alpha,\dot\delta)$ estimated at
some reference time, \eg $t_2$, by a fit to the observations. It is
possible to complete an attributable to a set of orbital elements by
adding the values of range and range rate $(\rho_2,\dot\rho_2)$ at the
same time\footnote{The epoch time of the elements is
$t_0=t_2-\rho_2/c$; these elements are said to be in Attributable
Coordinates \cite{ons2}.}. For each attributable we can determine an
\emph{Admissible Region} which is a compact set in the
$(\rho_2,\dot\rho_2)$ plane compatible with Solar System orbits
\cite{ons1}.

The optional intermediate steps are
\begin{enumerate}
\item an iterative \emph{Gauss map} to improve 
the solution of the degree 8 equation;
\item adding to the preliminary orbit(s) another one obtained from the
Attributable and a value for $(\rho_2,\dot\rho_2)$ selected inside the
\emph{Admissible region} (see details below);
\item a fit of the available observations to a 4-parameter
attributable at time $t_2$; the values of $\rho_2$ and
$\dot\rho_2$ are kept fixed at the previous values;
\item a fit of the available observations constrained to the
\emph{Line Of Variations (LOV)}, a smooth line defined by minimization
on hyperplanes orthogonal to the weak direction of the normal matrix.
\end{enumerate}

Intermediate step 1 has been discussed in Section~\ref{s:implementation}.

For intermediate step 2 we need to distinguish two cases depending
upon the topology of the Admissible Region \cite{ons1}. If it has two
connected components (this occurs for distant objects
observed near opposition) we select the point which is the center of
symmetry of the connected component farthest from the observer. This
corresponds to an orbit with $e<1$ although it is not, in general, a
circular orbit which may be incompatible with the Attributable.

If the Admissible Region is connected then we select the point along
the symmetry line $\dot\rho_2=const$ at $0.8$ times the maximum
distance $\rho_2$ compatible with $e\leq 1$. This case always occurs near
quadrature; if the object is indeed distant, thus has a low proper
motion $\eta$, the selected point is also far.

In any case, the selected point $(\rho_2,\dot\rho_2)$ in the
admissible region completed with the Attributable provides an orbit
which is compatible with the given Attributable and belongs to the
Solar System; this is called a \emph{Virtual Asteroid VA)}
\cite{virtast}.  The VA method provides an additional preliminary
orbit.  This does not matter when there are already good preliminary
orbits computed with Gauss' method (with significant curvature). We
shall see in Section~\ref{s:smalltest} that for TNOs this additional
preliminary orbit is required in many cases, the majority of
cases near quadrature, because the curvature is hardly significant.

Intermediate Step 3 is essentially the method proposed by D. Tholen
and also available in his public domain software KNOBS. It has already
been tested in the context of a simulation of a next generation
survey in \cite{acm05}.

Intermediate step 4 is described in full in \cite{multsol}. Our
preferred options are to use either Cartesian Coordinates or
Attributable Elements scaled as described in \cite[Table 1]{multsol}.

The steps listed above are all optional and indeed it is possible to
compute good orbits in many cases without some of them. However, if
the goal is a very reliable algorithm it is necessary to use them with
a smart connecting logic. As an example, Step 1 is used in a first
iteration, can be omitted in a second one. Step 2 is essential for
distant objects. Step 3 is used whenever the curvature is not
significant \ie when the observed arc is of type 1 \cite{discovery}
which can be tested \eg by eq.~(\ref{signcurv}). Step 4 is important
for weakly determined orbits, otherwise the differential corrections
may diverge when starting from an initial guess with comparatively
large residuals. Even then, Step 4 may fail and, in turn, diverge
under differential corrections.  In this case the differential
corrections restart from the outcome of the previous step. This
connecting logic is an extension of the one presented in
\cite[Figure 5]{multsol}.

\section{Tests} 
\label{s:tests}

The tests we are using are obtained by running a simulation based upon
a \emph{Solar System Model} - a catalog of orbits of synthetic objects
as described in \cite{acm05}. Given an assumed observation scheduling
and instrument performance we compute the \emph{detections} of the
catalog objects above a threshold signal to noise ratio and record the
corresponding simulated astrometric observation including astrometric
error. In the simulations used here we have not included false
detections (not corresponding to any synthetic object).

Then we assemble into \emph{tracklets} the detections (from the same
observing night) which could belong to the same object. The tracklets
are assembled in \emph{tracks} from several distinct nights (in
this context, at least three nights are required). For these
simulations we have used the algorithms of \cite{kubica} to assemble
both tracklets and tracks. 

When the number density of detections per unit area is low both
tracklets and tracks are (almost always) \emph{true} \ie they contain
only detections of one and the same synthetic object. When the number
density is large, as expected from the next generation surveys, both
tracklets and tracks can be \emph{false} (containing detections
belonging to different objects).  This is why a track needs to be
confirmed by computing an orbit: first a preliminary orbit, then by
differential corrections another orbit which fits all the observations
in the least squares sense. The structure containing the track and the
derived orbit with the accessory data necessary for quality control
(covariance matrix, weights, residuals, statistical tests) is called
an \emph{identification} \cite{DES}.

The purpose of the tests is to measure the performance of the
algorithms described in this paper, according to the following criteria:
\begin{itemize}

\item{Efficiency E:} the fraction of true tracks for which
good preliminary/least squares orbits were calculated.

\item{Accuracy A:} the fraction of returned orbits that correspond to
true tracks.  I.e., the orbit computation should fail on false tracks
(either no preliminary orbit or no least squares orbit or residuals
too large) .

\item{Speed S:}  Reciprocal of the CPU processing time.

\item{Goodness G:} the fraction of least squares orbit
close enough to the \emph{ground truth} orbit of the object to
 allow later recovery (\eg in another lunation).
\end{itemize}

The Speed criterion is less important than the others for
the reasons explained in Section~\ref{s:intro}. Nevertheless, we need
to check that the very large data sets expected from the next
generation surveys can be processed with
reasonable computational resources. 

Note that these tests should not be confused with tests of the
performance of next generation surveys like Pan-STARRS or LSST. The
purpose is to show that whatever the rate at which new objects are
observed their discoveries will not be lost because of inefficiency in
the orbit determination procedure.

\subsection{Small targeted test}
\label{s:smalltest}

Since the orbits of MBA and Jupiter Trojans are easier to compute than
NEOs and more distant objects \cite{acm05}, to assess the Efficiency
of these algorithms we have prepared four targeted simulations: two
containing only observations of NEOs and two with TNOs only. In both
cases, one simulation uses a surveying region near opposition and the
other surveys the so called \emph{sweet spots} at solar elongations
between $60^\circ$ and $90^\circ$.  The metrics being measured by
these tests are Efficiency and Goodness: Speed is irrelevant for such
small data sets and Accuracy is not a serious issue because the number
density per unit area is small (indeed, Accuracy is $100\%$ in all
tests of this Subsection).
\begin{table}[h]
\caption{For each of the NEO simulations, separately for objects
observed on a different number of nights, the columns give: [1] Total
number of objects, [2] Number of Complete Identifications
(containing all the tracklets belonging to the object), [3] Efficiency
(defined as [2]/[1], in $\%$), [4] Number of incomplete
Identifications, [5] Fraction [4]/[1] in $\%$ of incomplete
Identifications, [6] Number of objects
lost (no confirmed Identification), [7] Fraction [6]/[1] in $\%$ lost.}
\begin{center}
\label{t:neo}
\begin{tabular}{lrrrrrrr}
\hline
\hline\multicolumn{1}{c}{} & 
\multicolumn{1}{c}{[1]} & 
\multicolumn{1}{c}{[2]} &
\multicolumn{1}{c}{[3]} &
\multicolumn{1}{c}{[4]} &
\multicolumn{1}{c}{[5]} &
\multicolumn{1}{c}{[6]} &
\multicolumn{1}{c}{[7]}\\[1mm]
\multicolumn{1}{c}{Observed} & 
\multicolumn{1}{c}{} & 
\multicolumn{1}{c}{} &
\multicolumn{1}{c}{} &
\multicolumn{1}{c}{} &
\multicolumn{1}{c}{Inc.} &
\multicolumn{1}{c}{} &
\multicolumn{1}{c}{Lost}\\[1mm]
\multicolumn{1}{c}{Arc} & 
\multicolumn{1}{c}{Total} & 
\multicolumn{1}{c}{Compl.} &
\multicolumn{1}{c}{Effic.} &
\multicolumn{1}{c}{Inc.} &
\multicolumn{1}{c}{Fraction} &
\multicolumn{1}{c}{Lost} &
\multicolumn{1}{c}{Fraction}\\[1mm]
\hline
Opposition\\
3-nighters & 1123&  1119&   99.6\%&   0&    0.0\%&  4&    0.4\%\\
4-nighters &    2&     1&   50.0\%&   1&   50.0\%&  0&    0.0\%\\
6-nighters &  123&     0&    0.0\%& 123&  100.0\%&  0&    0.0\%\\[1mm]
\hline
Sweet Spots&\\
3-nighters &  397&   389&   98.0\%&   0&    0.0\%&  8&    2.0\%\\
6-nighters &   63&     0&    0.0\%&  63&  100.0\%&  0&    0.0\%\\[1mm]
\hline
\hline
\end{tabular}
\end{center}
\end{table}

The first part of Table~\ref{t:neo} refers to the simulation including
only NEO around opposition. Note that the Incomplete Identifications
are due to the fact that the algorithm used to assemble the tracks
operates on observations belonging to the same
lunation\footnote{An arc of one month has an excessive
curvature \cite[Section 8.2]{kubica}.} while the
simulation included two lunations. The separate orbits obtained in the
two lunations for the same object can be joined later with other
algorithms, e.g., the ones of \cite{ident4}. Thus there are
only 4 ``failures'',  true tracks which have not been confirmed
by the orbit computation. 
%
%Additionally, there was one case of a
%6-nighter for which two 3-night tracks were proposed but only
%one was confirmed. All these bad cases are discussed below.
%
The lower part of Table~\ref{t:neo} refers to the simulation including
only NEOs at the sweet spots. There are only 8 3-nighters without an
orbit.

% and three 6-nighters where one of the two 3-night tracks proposed
% did not lead to an orbit (the other incomplete identifications have
% two confirmed 3-night tracks for a 6-nighter due to the ``other
% lunation'' effect discussed above).

In conclusion, in both NEO simulations the Efficiency is very high but
not perfect, especially at the sweet spots. Note that in such small
tests it is always possible to increase the Efficiency to $100\%$ by
running additional iterations with increased computational
intensity. However, this would not provide useful indications on what
should be done with a much larger data set (e.g., we could increase
Efficiency at the expense of Accuracy).  Nevertheless, it is useful to
examine the few failure cases in order to learn about either the
limitations of our theory or defects in our implementation. None of
the cases in which an orbit was not computed, although a true track
was proposed, result from the failure of the differential corrections.
They all resulted from a failure of the preliminary orbit in the sense
discussed in Section~\ref{s:pretolsq}, in most cases because the
degree 8 equation has only invalid roots (either spurious, \ie
$\rho\leq 0$, or $\rho$ positive but small, resulting in a poor 2-body
approximation); the VA method was not of any help, as expected since
it is intended for low curvature cases. In other cases some
preliminary orbit could be found but the RMS of the fit was
comparatively large, in the range between $200$ and $300$ arcsec.

\begin{table}[h]
\caption{For each of the TNO simulations the same data as in
Table~\ref{t:neo}.}
\begin{center}
\label{t:tno}
\begin{tabular}{lrrrrrrr}
\hline
\hline\multicolumn{1}{c}{} & 
\multicolumn{1}{c}{[1]} & 
\multicolumn{1}{c}{[2]} &
\multicolumn{1}{c}{[3]} &
\multicolumn{1}{c}{[4]} &
\multicolumn{1}{c}{[5]} &
\multicolumn{1}{c}{[6]} &
\multicolumn{1}{c}{[7]}\\[1mm]
\multicolumn{1}{c}{Observed} & 
\multicolumn{1}{c}{} & 
\multicolumn{1}{c}{} &
\multicolumn{1}{c}{} &
\multicolumn{1}{c}{} &
\multicolumn{1}{c}{Inc.} &
\multicolumn{1}{c}{} &
\multicolumn{1}{c}{Lost}\\[1mm]
\multicolumn{1}{c}{Arc} & 
\multicolumn{1}{c}{Total} & 
\multicolumn{1}{c}{Compl.} &
\multicolumn{1}{c}{Effic.} &
\multicolumn{1}{c}{Inc.} &
\multicolumn{1}{c}{Fraction} &
\multicolumn{1}{c}{Lost} &
\multicolumn{1}{c}{Fraction}\\[1mm]
\hline
Opposition& \\
3-nighters & 2005&  2001&   99.8\%&   3&    0.15\%&   1&   0.05\%\\
4-nighters &   18&    18&  100.0\%&   0&    0.00\%&   0&   0.00\%\\
5-nighters &    3&     3&  100.0\%&   0&    0.00\%&   0&   0.00\%\\
6-nighters &  670&     0&    0.0\%& 670&   100.0\%&   0&   0.00\%\\
7-nighters &   13&     0&    0.0\%&  13&   100.0\%&   0&   0.00\%\\
8-nighters &    1&     0&    0.0\%&   1&   100.0\%&   0&   0.00\%\\
9-nighters &    6&     0&    0.0\%&   6&   100.0\%&   0&   0.00\%\\[1mm]
\hline
Sweet Spots & \\
3-nighters & 2493&  2491&   99.9\%&   0&    0.00\%&   2&   0.08\%\\[1mm]
\hline
\hline
\end{tabular}
\end{center}  
\end{table}

The upper part of Table~\ref{t:tno} refers to the simulation with TNOs
around opposition. The Incomplete identifications for 3-nighters are
cases in which more than one tracklet was available on some night. For
the cases for $\geq 6$ nights the track was not proposed for the same
reasons discussed in the NEO case above. Thus the orbit determination
has failed only in one case.  Moreover, this single failure is not due
to the preliminary orbits. The differential corrections stage computed
a nominal least square orbit that was then refused by the quality
control stage not because of small residuals (RMS $0.084$ arcsec) but
due to systematic trends (e.g., a slope) with a signal to noise
$\simeq 2.5$.

The lower part of Table~\ref{t:tno} refers to the simulation including
only TNOs in the sweet spot regions. Here the simulated scheduling included
only 3 nights and there are only two cases in which an orbit was not
computed. As in the opposition simulations these failures are due to
tight quality control thresholds because of systematic trends with signal to
noise between $3$ and $5.5$. 
\begin{table}[h]
\begin{center}
\caption{Fraction of objects lost 3-nighters using different algorithms.}
\label{t:method}
\begin{tabular}{lrrrrrrr}
\hline
\multicolumn{1}{c}{Simulation} & 
\multicolumn{1}{c}{Best} & 
\multicolumn{1}{c}{1 Pre} &
\multicolumn{1}{c}{1st It.} &
\multicolumn{1}{c}{No VA} &
\multicolumn{1}{c}{No 4fit} &
\multicolumn{1}{c}{No LOV} &
\multicolumn{1}{c}{LSQ}\\[1mm]
\hline
NEO Opp. & 0.40\%& 0.50\%&  3.2\%&   0.4\%&  0.4\%&    0.4\%&    0.4\%\\
NEO Sw.  & 2.00\%& 13.4\%& 18.1\%&   2.3\%&  2.3\%&    2.8\%&    2.8\%\\
TNO Opp. & 0.05\%& 0.05\%& 0.05\%&  38.3\%& 38.4\%&   45.7\%&   47.7\%\\
TNO Sw.  & 0.10\%& 0.10\%& 0.10\%&  75.3\%& 75.3\%&   82.6\%&   82.8\%\\[1mm]
\hline
\hline
\end{tabular}
\end{center}
\end{table}

In conclusion, the preliminary orbit algorithms have not shown even
one case of failure in the TNO simulations. However, we need to assess
the proportion of this success due to the improved but classical
method of Gauss rather than to the Virtual Asteroid method.  The
latter is expected to be especially effective for the low curvatures
typical of TNOs.  We have thus run again the simulations with a
version of the software not containing the VA method. The difference
of the results with the previous ones measures the contribution from
the VA method as shown in Table~\ref{t:method} in the column labeled
``No VA''.  In the opposition simulations giving up the VA method
resulted in $>1/3$ of the 3-night TNOs being lost due to a lack of
orbit computation. In the sweet spot simulations the same situation
occurred for $3/4$ of the 3-night TNOs.  The reason for this
difference is easily understood as near quadrature the TNO have an
even smaller proper motion than at opposition and the curvature is
very often not well measured. The conclusion is that the VA
method is essential for TNOs while it is almost irrelevant for NEO
(only a small contribution in the sweet spots).

For NEOs the most relevant question is the utility of all the care
we have exercised in ensuring that no identification is lost because of
double (or even triple) solutions of the preliminary orbit equations.
For this we have run a simulation in which only one preliminary orbit
was passed to differential corrections independent from the number
of solutions to the degree 8 equation. The selected preliminary
orbit was the one with lowest RMS of residuals. The results are in
Table~\ref{t:method} in the column ``1 Pre'' which clearly show that
to pass to differential corrections double (or possibly triple) solutions
near quadrature is essential for top Efficiency for NEOs. At opposition it
matters only in rare cases\footnote{The first example of
Section~\ref{s:examples} is the only case in which a NEO at opposition
would be lost by using only the preliminary orbit with lowest RMS as
shown by the small change in column ``1 Pre''.}.

Another test has been to stop after the first of the two iterations (see
Section~\ref{s:pretolsq}), the one with a tighter control in the RMS of the
residuals for the 2-body preliminary orbit (set at
$10$ arcsec in these tests) and using the Gauss map. From the results (column
``1st It.'') it is clear that the second iteration (with RMS of
preliminary orbit up to $100$ arcsec and the solution of degree 8
equation directly passed to differential corrections) has no effect at
all on TNOs but is relevant for NEOs especially in the sweet
spots. This is because the NEOs observed over 3 well spaced
nights are \emph{observed arcs of type 3} \cite{discovery}. \ie
the information contained in the observations is much more than just
$(\kappa,\dot\eta)$ of the entire arc which is used in the
preliminary orbit. On the other hand, the planetary perturbations
neglected in the preliminary orbit algorithms are large with respect to the 
observation accuracy (assumed to be $0.1$ arcsec). 

Although this paper is mostly about preliminary orbit algorithms we
need to also assess how much the improved differential corrections
algorithms (discussed in Section~\ref{s:pretolsq}) have contributed to
the overall success of these simulations. Column ``No 4fit'' reports
the results when the 4-parameter fit step was not used. The results
are essentially identical to the ones of the ``No VA'' case indicating
that the two algorithms must be used together for TNOs.

The column ``No LOV'' indicates that the step with the 5-parameter least
squares fit to obtain a LOV solution has a very large effect for TNO:
almost $1/2$ of the identifications in the Opposition case and $>4/5$
in the Sweet Spots case would not be confirmed by a least squares
orbit without the LOV solution. Indeed, the LOV
solution is the only one available for $53.8\%$ of the
TNOs near opposition and $81.6\%$ of the TNOs in the sweet spots (as
opposed to only $0.3\%$ of NEO at sweet spot and $0.1\%$ at
opposition). This is not a surprise. Indeed, the 3-night TNOs are almost
always observed arcs of type 2 (in some cases even type 1) which,
according to the tests in \cite{discovery}, are generally not suitable
to obtain a well determined orbit. We have also used this test to
confirm our statement that the \emph{weak direction} of the LOV is not
aligned in one special direction. In the opposition test the LOV has
an angle (computed with the proper scaling as indicated by
eq.~(\ref{covrrd})) between $-31^\circ$ and $+17^\circ$ with the
$\rho$ axis. In the sweet spots test it has an angle between
$-54^\circ$ and $+36^\circ$ with the $\dot\rho$ axis. In conclusion
the weak direction depends upon the elongation. It is closer to one of
the two axes but a different one in each of the two
cases\footnote{\cite{bernstein} warn that their arguments are not
applicable exactly at opposition and this is confirmed by our numerical
results near opposition.}.

We get the worst results (column 'LSQ') if neither the 4-fit algorithm
or the LOV solutions are used and the output of the preliminary orbit
is passed directly to a full 6-parameters differential corrections.
The difference between this column and the one labeled ``best''
measures the progress between using only the classical algorithms and
the best combination of algorithms we have found so far: for TNOs the
difference is very important, with all the steps discussed in
Section~\ref{s:pretolsq} essential to achieving the best results.

\subsection{Large scale tests}
\label{s:bigtest}

The main purpose of a large scale test is to measure the
Accuracy. Speed is not the limiting factor; besides, the orbit
determinations algorithms can be easily adapted for parallel
processing. Efficiency is not a problem for the overwhelming majority
of objects, which are MBA. Accuracy can affect Efficency: when there
are \emph{Discordant} identifications (with some tracklets in common)
if they cannot be merged (in an identification with all the tracklets
of both) there is no way to choose which of the two is the true
one. To keep the Accuracy high we then have to discard both, which
means losing true identifications and decreasing Efficiency. On the
other hand, we cannot afford low Accuracy since each false
identification introduces permanent damage to the quality of the
results\footnote{\emph{False facts are highly injurious to the
progress of science, for they often endure long...}, C. Darwin,
\emph{The Origin of Man}, 1871.}

We have prepared simulations for one lunation of a next generation
survey both near opposition and the sweet spots.  The limiting
magnitude was assumed to be V=24 and the Solar System model was used
at full density including the overwhelming majority of MBAs and
Trojans.  Table~\ref{t:simdata} gives the size of the dataset. The
focus of this paper is on the objects for which tracklets are
available in three nights.  Objects observed for a smaller number of
nights are part of the problem in that their tracklets can be
incorrectly identified: for such large number densities (per unit area
on the celestial sphere) false identifications happen easily
\cite[figure 3]{acm05}.  

\begin{table}[h]
\begin{center}
\caption{Simulated data sets: [1] survey region, [2] number of
tracklets, [3] of which false, [4] number of simulated objects with
observed tracklets, [5] of which with tracklets in 3 different nights,
[6] overhead (see text), [7] objects with tracklets in 2 different nights,
[8] with tracklets in only 1 night.}
\label{t:simdata}
\begin{tabular}{lccccccc} 
\hline
\multicolumn{1}{c}{[1]} & 
\multicolumn{1}{c}{[2]} & 
\multicolumn{1}{c}{[3]} &
\multicolumn{1}{c}{[4]} &
\multicolumn{1}{c}{[5]} &
\multicolumn{1}{c}{[6]} &
\multicolumn{1}{c}{[7]} &
\multicolumn{1}{c}{[8]}\\[1mm]
\hline
Oppos.&  654315& 26006& 253289& 164333& 222.8& 41244&  47712\\
Sweet sp.& 695067& 59253& 283831& 144903& 501.3& 62177&  76751\\[1mm]
\hline
\hline
\end{tabular}
\end{center}
\end{table}

The first problem concerning Accuracy occurs at the tracklet
composition stage: some tracklets are false, that is, they mix
detections belonging to different objects. The question is whether
they are identified, thus decreasing Accuracy.

The second Accuracy problem occurs at the track composition stage. A
track is just a \emph{hypothesis of identification} to be checked by
computing an orbit: at a high tracklet number density most of the
tracks are false. The \emph{Overhead} (column marked \emph{Overh.}) is
the ratio between the total number of proposed tracks and true ones:
it was large, at the sweet spots even above what was found in previous
simulations \cite[Table 3]{kubica}.

The question is whether the orbit determination stage can produce the
true orbits with good Efficiency and still reject almost all the false
tracks. To achieve this, the residuals of the best fit orbits need to
be submitted to a rigorous statistical quality
control\footnote{This quality control is currently done with the
intervention of a human eye looking at the residuals. However, given
the number of orbits, the quality control needs to be
fully automatized.}. Our residuals quality control algorithm uses the
following 10 metrics (control values in square brackets)
\begin{itemize}
\item RMS of astrometric residuals divided by the assumed RMS of the
observation errors (=$0.1$ arcsec in these simulations) [$1.0$] 
\item RMS of photometric residuals in magnitudes [$0.5$]
\item bias of the residuals in RA and in DEC [$1.5$]
\item first derivative of the residuals in RA and in DEC [$1.5$]
\item second derivative of the residuals in RA and in DEC [$1.5$]
\item third derivative of the residuals in RA and in DEC [$1.5$]
\end{itemize}
To compute the bias and derivatives of the residuals we fit them to a
polynomial of degree 3 and divide the coefficients by their standard
deviation as obtained from the covariance matrix of the
fit\footnote{When these algorithms are used on real data
additional metrics should take into account the outcome of
outlier removal \cite{carpino}. For simulations this does not apply.}.

\begin{table}[h]
\begin{center}
\caption{Accuracy Results: before and after normalization, total
number of false identifications accepted, percentage (with respect to
the total number of identifications), number of identifications
containing false tracklets.}
\label{t:accu}
\begin{tabular}{lrrrrrr} 
\hline
\multicolumn{1}{c}{Region} & 
\multicolumn{3}{c}{All Identifications}&
\multicolumn{3}{c}{Normalized}\\
&\multicolumn{1}{c}{False} & 
\multicolumn{1}{c}{\%} &
\multicolumn{1}{c}{F.Tr.} &
\multicolumn{1}{c}{False} & 
\multicolumn{1}{c}{\%} &
\multicolumn{1}{c}{F.Tr.}\\[1mm]
\hline
Oppos.   &  7093&  4.31&   4&      80& 0.05&  1\\
Sweet sp.& 1869&   1.30&  10&      29& 0.02&  0\\[1mm]
\hline
\hline
\end{tabular}
\end{center}
\end{table}

The results are summarized in Tables~\ref{t:accu} and \ref{t:effi}.
As expected, the main problem is in Accuracy. Notwithstanding the
tight statistical quality controls on residuals, while processing tens
of millions of proposed tracks a few thousands false tracks are found
to fit well all their tracklets (columns marked \emph{False}).  The
numbers are very small with respect to the total number of tracks but
they are not negligible with respect to the number of true tracks
(columns marked \emph{\%}). This happens by combining tracklets from 2
(or 3) distinct simulated objects. A much smaller number of false
tracks contain some false tracklets (columns marked \emph{F.Tr.}).
Thus, even the presence of a significant fraction of false tracklets
does affect neither Efficiency nor Accuracy.

Given cases like these, with a fit passing all the quality controls,
we cannot a priori discard any of them since only by consulting the
\emph{ground truth} can we know they are \emph{false}. By further
tightening the quality control parameters we may remove many false
true but some true identifications as well. The values of the controls
used are already the result of adjustment suggested by experiments to
find a good trade off between Accuracy and Efficiency.

The most effective method to remove false tracks is obtained by not
considering each identification by itself but globally. We have
previously defined the \emph{normalization} of lists of
identifications in \cite[Section 7]{ons2} and \cite[Section
6]{acm05}. It removes duplications and inferior identifications but
also rejects all the Discordant identifications. This is not because
they are all presumed false, indeed very often one true and one false
identification are Discordant, but we do not know which is
which. Thus, according to our philosophy giving paramount importance
to the reliability of the results we remove both and sacrifice
Efficiency for Accuracy. We have refined the normalization procedure
by checking, for Discordant identification, whether one of the two is
significantly superior to the other, by comparing the normalized RMS
of the astrometric residuals: if the difference is more than $0.25$ we
keep the best. The results of this normalization procedure are shown
on the right hand side of table~\ref{t:accu}. It is clear that the
number of false tracks can be reduced to negligible values in this
way.

\begin{table}[h]
\begin{center}
\caption{Efficiency Results}
\label{t:effi}
\begin{tabular}{lrccrcc} 
\hline
\multicolumn{1}{c}{Obj.Type} & 
\multicolumn{3}{c}{Opposition}&
\multicolumn{3}{c}{Sweet Spots}\\
&\multicolumn{1}{c}{Total} & 
\multicolumn{1}{c}{Eff.\%} &
\multicolumn{1}{c}{Eff.No.\%} &
\multicolumn{1}{c}{Total} & 
\multicolumn{1}{c}{Eff.\%} &
\multicolumn{1}{c}{Eff.No.\%}\\[1mm]
\hline
All&       161146&  97.3&   95.9&    144903&   98.0&  97.4\\
MBA&       154700&  97.3&   95.8&    135911&   98.0&  97.4\\
NEO&          353&  90.4&   90.4&       271&   80.1&  80.1\\
Tro&             &      &       &      6894&   97.9&  97.8\\
Com&          665&  98.6&   97.6&       253&   98.0&  97.6\\
TNO&         5428&  97.7&   97.7&      1574&   98.7&  98.7\\[1mm]
\hline
\hline
\end{tabular}
\end{center}
\end{table}

However, the Efficiency also changes as a result of the normalization.
In Table~\ref{t:effi} the rows indicate how the results differ
according to the orbital class of the simulated objects. The row
marked \emph{Com} includes Centaurs, long period and short period
comets.  Note that the sweet spots simulation does not detect any
Jupiter Trojans because the Trojan swarms were not in these directions
at the time of the synthetic observations.  The columns marked
\emph{Eff.\%} show the Efficiency before and after \emph{Eff.No.\%}
normalization. As a rule of thumb, on average a little more than 1
true identification is lost for each false rejected.

Table~\ref{t:effi} refers to the objects observed with 3 tracklets in
3 nights. A minor additional problem occurs in the opposition
simulation. For the $3,187$ objects observed with $>3$ tracklets in 3
nights the proposed tracks may be true but \emph{incomplete}, e.g.,
have only 3 tracklets when the corresponding object has more.  In
fact, for $13.5\%$ of these objects one or more tracklets fail to be
included in the identification\footnote{This incompleteness was in
fact a problem of interface between two processing steps: complete
identifications could have been found by the track composition
algorithms.}. Although their number is not large these incomplete
identifications are difficult to be fixed at some later processing
stage. The solution is to run an algorithm of \emph{attribution}
\cite{ident4} to add the omitted tracklets. We have tested this method
and found it to be $99.8\%$ efficient.

The question then is: are the results of Tables~\ref{t:accu} and
\ref{t:effi} satisfactory and, if not, what else can be done?  Note
that the Efficiency for NEOs and TNOs is not affected by normalization
(because the number densities are much lower). It might be argued that
loosing a few percent of MBAs is not important. Nevertheless, we claim
that even this problem can be solved together with the other, possibly
more important, of tens of NEOs and comets lost for other
reasons.

By separately analyzing the Efficiency of the three steps of the
procedure (track composition, orbit computation, normalization) we
have established that the algorithm to generate tracks has been
$97.6\%$ efficient at opposition and $98.7\%$ at the sweet spots. The
efficiency of the orbit computation procedure on the proposed true
tracks has been $99.8\%$ efficient at opposition and $99.3\%$ at the
sweet spots. The normalization procedure has been $98.6\%$ efficient
at opposition and $99.4\%$ at the sweet spots. It follows that the
different steps are well balanced in their performances and there is
not much room for improvement. Thus the solution is to use a two
iteration procedure.

The normalization procedure generates two outputs: the new list of
identifications and the list of \emph{leftover} tracklets which have
not been used in the confirmed identifications. Note that when two
tracklets are \emph{Discordant} (have detections in common) if one of
the two is included in one confirmed identification then the other can
be considered \emph{used}. In this way the set of tracklets is reduced
and many false tracklets are discarded. In the opposition simulation
there are $168,122$ leftover tracklets of which $5,363$ are false: a
reduction by $74.3\%$ of the original dataset and a reduction by
$79.4\%$ in the number of false tracklets. At the sweet spots the
corresponding numbers are $232,101$ leftover tracklets of which
$17,033$ are false: a reduction by $66.6\%$ and $71.2\%$ respectively.

The leftover tracklets can be used as input to another iteration which
could use the same algorithms as the first one (maybe with different
controls and options) or could use very different methods; because of
the reduced number density of tracklets, Accuracy should be less of a
problem.  
One possibility is to iterate the same procedure: generate tracks
starting from the leftover tracklets with the same algorithm
\cite{kubica} but using different options; then run again the orbit
determination, possibly with different options, and the
normalization. Another possibility is to use for a second iteration a
completely independent algorithm to find identifications. Such an
algorithm has already been proposed and tested on large simulations
\cite{ons2} and on small real data sets \cite{boattini}. 

A full discussion of the iteration strategies to be used for the next
generation surveys is beyond the scope of this paper.
However, to show that the normalized Efficiency values shown in
Table~\ref{t:effi} are not to be considered a problem, we have run an
improved version of the recursive attribution algorithms of
\cite{ons2} on the leftover tracklets\footnote{To better control the
false identifications we have used even tighter quality
controls.}. The results for the opposition simulation are as follows:
we have been able to recover $75.4\%$ of the lost objects ($85.3\%$ of
the lost NEOs), thus bringing the overall Efficiency to $99.0\%$
($97.1\%$ for NEOs); for the sweet spots the values are $75.0\%$
recovery ($85.2\%$ for NEOs) and $99.4\%$ overall Efficiency ($97.1\%$
for NEOs). The false identifications remain very few even
including the second iteration results: only 86 ($0.06\%$) at
opposition and only 29 ($0.02\%$) at the sweet spots. The same
procedure allows also  to compute normalized identifications
for 2-nighters, with Efficiency $83.4\%$ and only $2.1\%$ false
identifications at opposition.  The corresponding values in the sweet
spots are $89.2\%$ Efficiency and $1.3\%$ false identifications.

The best way to assess Goodness of the results is to try to perform a
simulation mimicking as much as possible the way in which the results
from one lunation would be used in the next one. Thus we have made two
complete simulations at opposition for two consecutive lunations.
Having obtained 3-nighter identifications for the first month we have
attempted to attribute to each one of them some tracklets in the other
month. Using the same quality controls as for orbit determination
within the same lunation this procedure was $99.6\%$ efficient for
objects with 1 tracklet in the second lunation, $99.7\%$ efficient for
objects with 2 tracklets and $99.9\%$ efficient for objects with 3
tracklets. There were no NEOs among the few cases of either failed or
incomplete attribution.

\section{Conclusions and Future Work}

The purpose of this paper was to identify efficient algorithms to
compute preliminary and least squares orbits given a track
(or proposed identification).

We have found suitable algorithms by revising the classical
\emph{preliminary orbit} methods. The most important improvements are
provisions to keep alternate solutions under control. The
existence of double solutions was known since long time and we have shown
that even triple solutions can occur.  Still there is no reason this
should impair the performance of the orbit determination algorithms, provided
these cases are handled with due care.

For the \emph{differential corrections} stage, leading from
preliminary orbits to least squares ones, we have adopted algorithms
innovative with respect to the classical ones but already available
from our previous work. If properly combined with a good control logic
they significantly improve the efficiency of differential corrections
even when the preliminary orbits are not close to the nominal solution
(e.g., because of low curvature).

The third stage of orbit determination is the \emph{quality control}
of the results imposed by applying statistical criteria to the
residuals of the least squares fit. When there is just one object (or
only a small number) under consideration at a time this stage may
appear unnecessary. However, with the high sky-plane detection density
to be expected with the next generation surveys this stage turns out
to be the most critical one. When the number density of observations
(per unit area on the sky) is large tracklets belonging to different
objects may be incorrectly identified. To clean up the output from
these false identifications is not easy.  We have found the 
method of \emph{normalization} to be very effective in removing false
identifications, but unavoidably some true identifications are
sacrificed to remove the discordant false ones. The critical point is
to select options and details of the algorithms in such a way that the
number of false identifications is kept to a very low value but few
true ones are lost.

Although our mathematically rigorous theoretical results do not need
confirmation it has been useful to test their practical performance on
simulations of the next generation surveys. In this way we have shown
that orbits can be computed even for the most difficult classes of
orbits. We have also shown, with full density simulations including an
overwhelming majority of MBA, that the large number of objects
observed does not result in a ``false identification catastrophe''. On
the contrary, a large number density is compatible with a low number
of lost objects provided the quality control on the residuals is tight
enough and the sequence of algorithms is suitably chosen. 

%When the same methods will be applied to large real datasets, the
%almost perfect performances of the simulations are not likely to be
%possible and anyway could not be verified.

The performance of a procedure for identification and orbit
determination critically depends upon the performance of the
individual algorithms and upon the \emph{pipeline design} - the
sequence of algorithms operating one upon the output of another. We
have used the algorithms from \cite{kubica} as the first step,
followed by the algorithms of this paper as the second step. We have
mentioned the possibility of using the algorithms of \cite{ons2} as the
third step. Even more complicated pipelines can be conceived and may
be superior in performance. However, a detailed discussion of pipeline
design is beyond the scope of this paper and will be the subject of
future work.

Another subject of future work is the definition of the procedure to
combine the results from different lunations of a large survey.  We
have done a small test with a second lunation by using the
\emph{attribution} algorithm of \cite{ident4}. Under other conditions,
when a survey has been operating for more than one year, other
algorithms such as the ones of \cite{ident3} and a variant of the
methods of \cite{kubica} may become necessary.

In the three papers \cite{ons2}, \cite{kubica} and the present one we
have defined a number of algorithms to be used to process astrometric
data of Solar System objects when the number density will be much
larger than that currently observed.  This will very soon be the case
with the next generation surveys including Pan-STARRS and LSST. Such
algorithm definition work is a necessary step to exploit
their superior survey performance and provide identifications and
orbits for most observed objects. In the near future we will need to
handle real data which, of course, will contain unpredicted problems
and present a new, formidable challenge.

\section*{Acknowledgments}

Milani \& Gronchi are supported by the Italian Space Agency through
contract 2007-XXXX, Kne\v zevi\'c from Ministry of Science of Serbia
through project 146004 "Dynamics of Celestial Bodies, Systems and
Populations".  Jedicke \& Denneau are supported by the Panoramic
Survey Telescope and Rapid Response System at the University of
Hawaii's Institute for Astronomy funded by the United States Air Force
Research Laboratory (AFRL, Albuquerque, NM) through grant number
F29601-02-1-0268.  Pierfederici is supported by the LSST's research
and development effort funded in part by the National Science
Foundation under Scientific Program Order No. 9 (AST-0551161) through
Cooperative Agreement AST-0132798.  Additional LSST funding comes from
private donations, in-kind support at Department of Energy
laboratories and other LSSTC Institutional
Members.

\end{document}